\documentclass[preprint,12pt]{elsarticle}
\usepackage{framed,multirow}
\usepackage{array}
\newcolumntype{P}[1]{>{\centering\arraybackslash}p{#1}}
\usepackage{graphicx}
\usepackage{float}
\usepackage{changepage} 
\usepackage{booktabs, multirow}

\usepackage{subcaption,graphicx}
\usepackage{pdfpages}
\usepackage{tikz}
\usepackage{amsmath}
\usepackage[ruled,linesnumbered]{algorithm2e}
\usepackage{amssymb}
\usepackage{latexsym}
\usepackage{mathtools}
\usepackage{hyperref}
\usepackage{diagbox}
\usepackage{makecell}
\usepackage{float}
\hypersetup{
	colorlinks=true,
	linkcolor=blue,
	filecolor=magenta,
	urlcolor=cyan,
}
\usepackage{longtable}
\usepackage{bm}
\usepackage{url}
\usepackage{xcolor}
\usepackage{color} 
\usepackage[dvipsnames]{xcolor}
\definecolor{newcolor}{rgb}{.8,.349,.1}
\usepackage{geometry}

\usepackage{lineno}
\usepackage{caption}
\usepackage{printlen}
\usepackage[table]{xcolor}
\journal{Journal of computational physics}

\begin{document}

\begin{frontmatter}
\title{Discontinuity-aware KAN-based physics-informed neural networks}
\author[1,2]{Guoqiang Lei}
\author[1]{D. Exposito}
\author[3,1]{Xuerui Mao\corref{cor1}}
\cortext[cor1]{Corresponding author.}\ead{maoxuerui@sina.com}

\affiliation[1]{
	organization={School of Interdisciplinary Science, Beijing Institute of Technology},
	city={Beijing},
	postcode={100081}, 
	country={China}}
\affiliation[2]{
	organization={School of Mechatronical Engineering, Beijing Institute of Technology},
	city={Beijing},
	postcode={100081}, 
	country={China}}
\affiliation[3]{
	organization={Beijing Institute of Technology (Zhuhai)},
	city={Zhuhai},
	postcode={519088}, 
	country={China}}

\begin{abstract}
Physics-informed neural networks (PINNs) have proven to be a promising method for the rapid solving of partial differential equations (PDEs) in both forward and inverse problems. However, due to the smoothness assumption of functions approximated by general neural networks, PINNs are prone to spectral bias and numerical instability and suffer from reduced accuracy when solving PDEs with sharp spatial transitions or fast temporal evolution. To address this limitation, a discontinuity-aware physics-informed neural network (DPINN) method is proposed. It incorporates an adaptive Fourier-feature embedding layer to mitigate spectral bias and capture steep gradients, a discontinuity-aware network that generalizes the Kolmogorov representation theorem to the discontinuous regime for the modeling of shock-wave properties, mesh transformation to accelerate convergence across complex geometries, and learnable local artificial viscosity to stabilize the algorithm near discontinuities. In numerical experiments regarding the inviscid Burgers' equation, Riemann problems, and transonic and supersonic airfoil flows, DPINN demonstrated superior accuracy in capturing discontinuities compared to existing methods.
\end{abstract}
\end{frontmatter}
\section{Introduction}
\label{sub:1}
Partial differential equations (PDEs) with discontinuous solutions are prevalent in science and engineering applications, such as aerodynamics \cite{gaitonde2013progress}, astrophysical explosions \cite{longair2011high}, and multiphase flows \cite{jalili2024physics}. Solving these PDEs presents a significant numerical challenge. Conventional methods, such as spectral schemes \cite{gottlieb1981spectral}, struggle with steep gradients and suffer from the Gibbs phenomenon \cite{gottlieb1997gibbs}, resulting in non-physical oscillations near discontinuities. Over the past decades, numerous numerical techniques have been developed to stabilize solutions near discontinuities while maintaining accuracy in smooth regions. These methods include equation modification techniques, such as artificial viscosity (AV) methods \cite{vonneumann1950method}, which introduce dissipation to smooth discontinuities at the expense of accuracy, and high-resolution schemes, such as weighted essentially non-oscillatory (WENO) approaches \cite{liu1994weighted}. For a comprehensive overview of numerical methods regarding the calculation of discontinuities, see \cite{godunov1959finite,bassi1997high,pirozzoli2011numerical}. However, the high computational cost of these numerical methods limits their application in outer-loop tasks, such as optimization, design, control, data assimilation, and uncertainty quantification, which require multiple repetitions with different parameters \cite{kramer2024learning}.

Physics‐informed neural networks (PINNs) have emerged as a promising approach for the rapid solving of parameterized PDEs \cite{raissi2019physics}. These models encode physical laws into the loss function and seek solutions through the optimization of network parameters. However, conventional PINNs are unsuitable for solving PDEs with discontinuous solutions due to an optimization paradox near discontinuities, which induces oscillatory loss during training and prevents effective convergence, resulting in numerical instability \cite{liu2024discontinuity,wang2021understanding, rahaman2019spectral,wang2022and}. In recent years, numerous methodologies have been developed to overcome the limitations mentioned above, including physical modification, loss and training modification, and architecture modification methods \cite{abbasi2025challenges}.

Physical modification strategies fundamentally reformulate the physical problem by modifying the governing equations to mitigate the inherent challenge of gradient explosions at discontinuities. For instance, artificial viscosity methods introduce dissipation terms to smooth the sharp transitions, applied globally \cite{raissi2019physics} or locally \cite{wassing2025adopting,wassing2024physics} with fixed \cite{fuks2020limitations} or adaptive \cite{coutinho2023physics} intensity. Moreover, such strategies include total variation penalization \cite{patel2022thermodynamically}, entropy condition \cite{diab2022data,huang2023limitations,jagtap2022physics}, change of variables \cite{abbasi2023simulation,wu2024variable}, and weak‑form PINNs \cite{de2024wpinns,kharazmi2019variational,chaumet2022improving}. Loss and training modification methods modify the network loss landscape during training to promote convergence, such as weighted‑equations PINNs \cite{liu2024discontinuity,neelan2024physics}, gradient‑annihilated PINNs \cite{ferrer2024gradient}, advanced collocation‑point sampling \cite{yang2024moving,mao2020physics}, and higher‐order optimizers \cite{urban2025unveiling,tancik2020fourier,kiyani2025optimizer}. Architectural modification techniques, such as transformer‑based architectures \cite{rodriguez2022physics,anagnostopoulos2024residual}, extended PINNs \cite{jagtap2020extended}, Fourier-feature embeddings \cite{wang2021eigenvector,bruno2022fc}, Lagrangian PINNs \cite{mojgani2023kolmogorov}, and adaptive activation functions \cite{jagtap2020adaptive}, adjust the network structure to improve their ability to capture shock-wave features but often introduce more trainable parameters, increasing computational costs. Extensive investigations have demonstrated the ability of PINNs to resolve PDEs with discontinuous solutions. However, balancing computational efficiency, accuracy, and robustness remains a challenging task. PINNs have only recently been successfully applied to simulate transonic flow around an airfoil by introducing localized artificial viscosity \cite{wassing2024physics}, but only smooth shock-waves have been obtained, and their resolution remains insufficient.

In the present work, we develop a discontinuity-aware physics-informed neural network (DPINN) method that improves accuracy and significantly reduces the number of parameters required for discontinuity calculation compared to existing methods. This method is built on an adaptive Fourier-feature embedding layer and a discontinuity-aware Kolmogorov–Arnold network (DKAN), enabling the automatic detection and capturing of discontinuities. Additionally, mesh transformation and learnable local artificial viscosity strategies are further incorporated to accelerate convergence for complex geometries and stabilize solutions near discontinuities, respectively.

The remainder of this paper is organized as follows: Section \ref{sub:2} provides an overview of the DPINN method; Section \ref{sub:3} demonstrates the greater performance of the DPINN method compared to existing technologies through simulations of the Burgers' equation and transonic and supersonic flows around an airfoil; Section \ref{sub:4} concludes this study, discusses its contributions and limitations, and proposes future research directions.
\section{Methodology}
\label{sub:2}
\begin{figure}[!h]
	\centering
	\includegraphics{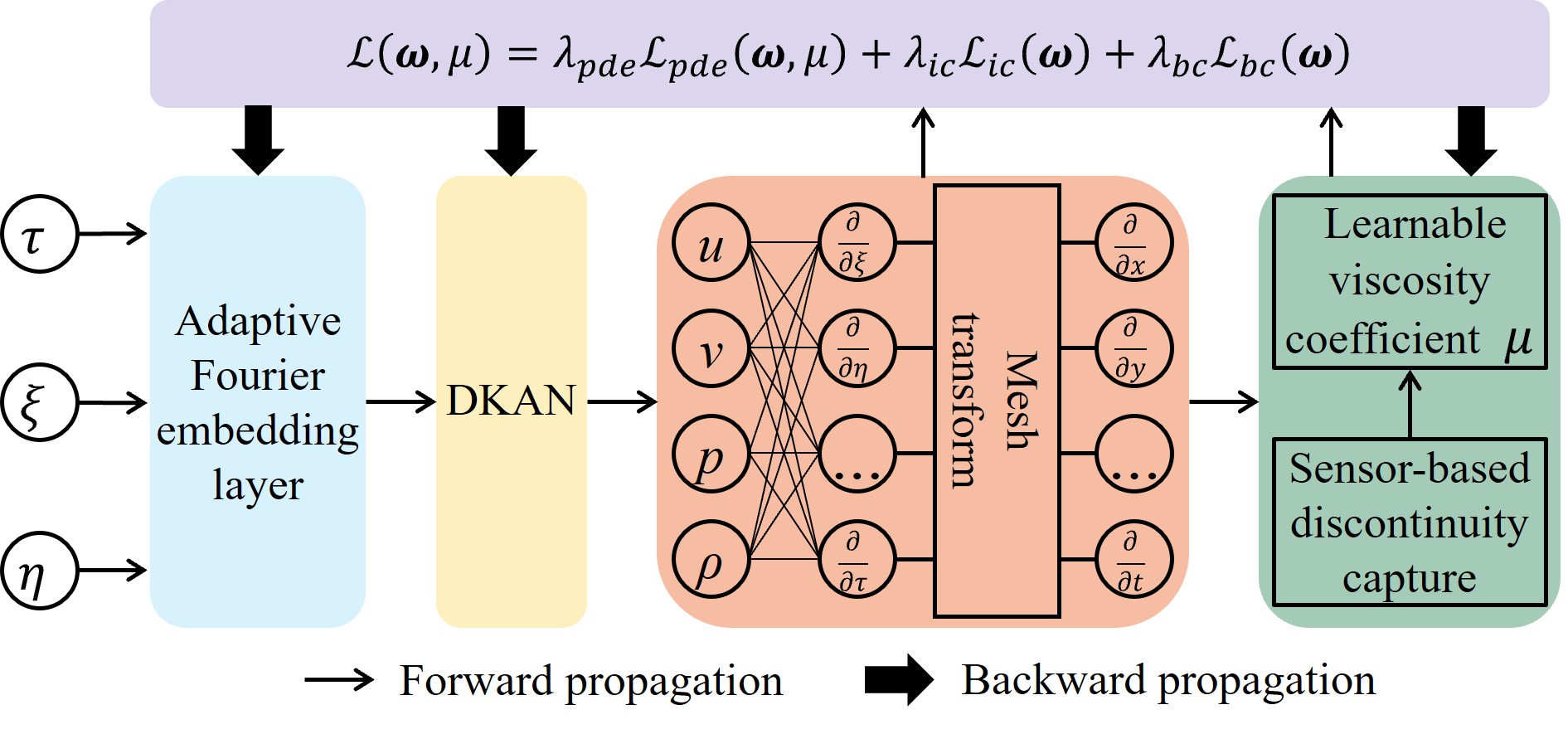}
	\caption{Flowchart of the DPINN method applied to a two-dimensional (2D) unsteady flow.}
	\label{fig: DPINN}
\end{figure}

As mentioned above, the DPINN method establishes an artificial intelligence (AI)-based solver particularly suitable for solving PDEs with discontinuous solutions, incorporating architecture modification, loss and training modification, and physics modification strategies, as shown in Fig. \ref{fig: DPINN}. First, the PINN architecture is used to solve PDEs without relying on prior information from the data. Second, the adaptive Fourier-feature embedding layer mitigates spectral bias and captures sharp changes. Third, the DKAN architecture, composed of discontinuity-aware activation functions, accurately models shock-wave features. Fourth, mesh transformation improves convergence for complex geometries. Finally, local learnable artificial viscosity, restricted to regions identified by a shock sensor, stabilizes solutions near discontinuities.

\subsection{Physics-informed neural network}
\label{sub:2.1}
In this section, we first provide a brief review of the PINN methodology for PDEs:

\begin{equation}
\begin{aligned}
	\mathcal{N}[\bm{u}(\bm{x}, t)]&=0, \bm{x} \in \Omega, t \in(0, T], \\
	\mathcal{I}[\bm{u}(\bm{x}, 0)]&=0, \bm{x} \in \Omega, \\
	\mathcal{B}[\bm{u}(\bm{x}, t)]&=0, \bm{x} \in \partial \Omega, t \in(0, T],
\end{aligned}
	\label{eq: pde}
\end{equation}
where $\bm{u}(\bm{x}, t)$ is the solution, and $\mathcal{N}[\cdot]$, $\mathcal{I}[\cdot]$, and $\mathcal{B}[\cdot]$ denote the PDE operator, initial condition operator, and boundary condition operator, respectively. $\Omega$ and $\partial \Omega$ represent the computational domain and its boundary, respectively, and $T$ is the final time. In contrast to numerical methods that solve PDEs iteratively, PINN takes the spatial coordinates $\bm{x} \in \Omega$ and time $t \in[0, T]$ as input and optimizes the neural network parameters $\bm{w}$ to minimize the loss function $\mathcal{L}(\bm{w})$, resulting in the optimal parameters $\bm{w}^*$ and approximate solutions:
\begin{equation}
	\begin{aligned}
		\bm{w}^*=\underset{\bm{w}}{\operatorname{argmin}} \mathcal{L}(\bm{w}) &= \underset{\bm{w}}{\operatorname{argmin}} \left( \lambda_{pde} \mathcal{L}_{pde} + \lambda_{ic} \mathcal{L}_{ic} + \lambda_{bc} \mathcal{L}_{bc} \right), \\
		\mathcal{L}_{pde} &= \| \mathcal{N}\left[\bm{u}(\bm{x}, t; \bm{w}) \right] \|^2_2, \bm{x} \in \Omega, t \in(0, T], \\
		\mathcal{L}_{ic} &= \| \mathcal{I}\left[\bm{u}(\bm{x}, 0 ; \bm{w})\right] \|^2_2, \bm{x} \in \Omega, \\
		\mathcal{L}_{bc} &= \| \mathcal{B}\left[\bm{u}(\bm{x}, t; \bm{w}) \right] \|^2_2,\bm{x} \in \partial \Omega, t \in(0, T],
	\end{aligned}
	\label{eq: loss}
\end{equation}
where $\left\|\cdot\right\|_2$ represents the $L_2$ norm, $\mathcal{L}_{pde}$, $\mathcal{L}_{ic}$ and $\mathcal{L}_{bc}$ denote the PDE, initial condition, and boundary condition losses, respectively, and the non-negative weights $\lambda_{pde}$, $\lambda_{ic}$ and $\lambda_{bc}$ control the trade-off between these components. Once the loss function drops to a sufficiently small value, the neural network solution satisfies the governing equation constraints approximately. Provided the solution is unique, the output of the neural network corresponds to the solution of the PDE system in Eq. \eqref{eq: pde} \cite{raissi2019physics}.

\subsection{Adaptive Fourier-feature embedding}
\label{sub:2.2}
Due to spectral bias, standard neural networks fail to effectively represent high-frequency components, which limits their applicability to multi-frequency problems \cite{tancik2020fourier,catalani2024neural}. Recently, the Random Fourier‐feature (RFF) embedding has been applied to mitigate this limitation by introducing a non-trainable Fourier layer \cite{wang2021eigenvector,bruno2022fc}, which encodes inputs using trigonometric functions with randomly sampled frequencies, thereby improving high-dimensional feature representation, defined as:

\begin{equation}
	\operatorname{RFF}(\bm{x})=\frac{1}{\sqrt{m}} \left[\sin \left(2 \pi \bm{B} \bm{x}\right), \cos \left(2 \pi \bm{B} \bm{x}\right)\right]^\top \in \mathbb{R}^{2m},
\end{equation}
where $\bm{x} \in \mathbb{R}^{n_{in}}$ is the input vector of dimension $n_{in}$, $\top$ denotes the matrix transpose, and $\bm{B} \in \mathbb{R}^{m \times n_{in}}$ comprises $m$ frequency components independently sampled from a zero-centered Gaussian distribution with standard deviation $\sigma$, controlling the frequency range of the encoded vector. A small value of $\sigma$ acts as a low-pass filter, leading to underfitting due to the failure to capture signal details, while a large value may introduce high-frequency noise, resulting in overfitting \cite{catalani2024neural}. For all Multilayer Perceptron (MLP)-based architectures, a value of $\sigma=1$ is employed herein.

\begin{figure}[!h]
	\centering
	\includegraphics{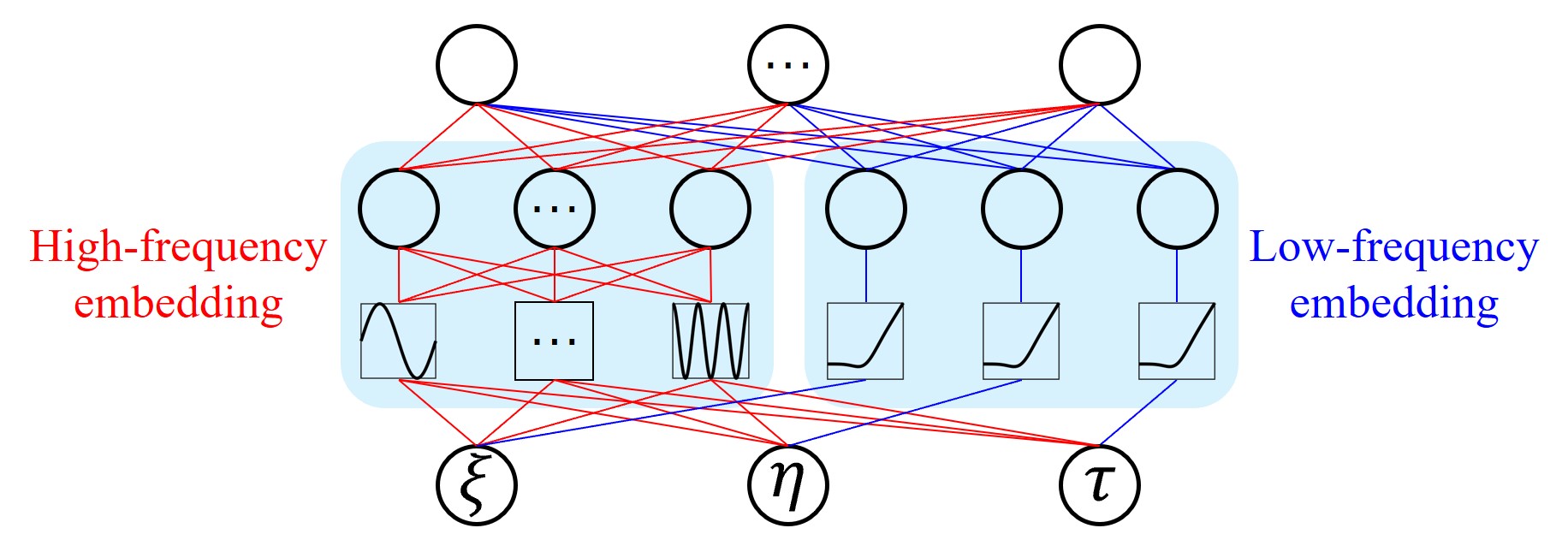}
	\caption{Schematic of the adaptive Fourier embedding layer with red and blue channels representing high- and low-frequency embeddings, respectively.}
	\label{fig: KAF}
\end{figure}

Tuning $\sigma$ to capture multi-frequency features accurately is time‐consuming. To overcome this limitation, a Kolmogorov-Arnold Fourier network (KAF) is introduced as shown in Fig. \ref{fig: KAF}, enabling frequency-adaptive Fourier-feature embedding through its learnable hybrid spectral correction mechanism \cite{zhang2025kolmogorov}, defined as:
\begin{equation}
	\operatorname{KAF}(\bm{x})=\underbrace{\bm{W}_l \operatorname{Gelu}(\bm{x})}_{\text{low frequency}}+ \underbrace{\bm{W}_h {\mathcal{F}}(\bm{x})}_{\text{high frequency}} \in \mathbb{R}^{n_{out}},
\end{equation}
where $\bm{W}_l \in \mathbb{R}^{n_{out} \times n_{in}}$ and $\bm{W}_h \in \mathbb{R}^{n_{out} \times 2m}$ denote trainable scaling matrices for low- and high-frequency components, respectively, and $n_{in}$, $n_{out}$, $m$ are the input dimension, output dimension, and number of frequency components. High-frequency weights $\bm{W}_h$ are initialized at a much smaller scale than low-frequency weights $\bm{W}_l$ to prioritize low-frequency feature learning. As training progresses, $\bm{W}_h$ increases to improve high-frequency representation. The functions $\operatorname{Gelu}(\bm{x})$, capturing low-frequency large-scale structures, and $\mathcal{F}(\bm{x})$, expanding spectral coverage, are defined as:
\begin{equation}
	\begin{aligned}
		\operatorname{Gelu}(\bm{x})&=0.5 \bm{x}\left(1+\tanh \left[\sqrt{\frac{2}{\pi}}\left(\bm{x}+0.044715 \bm{x}^3\right)\right]\right) \in \mathbb{R}^{n_{in}},\\
		{\mathcal{F}}(\bm{x})&={\frac{1}{\sqrt{m}}}\left[\sin \left(2 \pi \bm{B}_f \bm{x}\right), \cos \left(2 \pi \bm{B}_f \bm{x}\right)\right]^\top \in \mathbb{R}^{2m},
	\end{aligned}
\end{equation}
where $\bm{B}_f \in \mathbb{R}^{m \times n_{in}}$ is the learnable frequency matrix used to capture potential high-frequency discontinuous features \cite{zhang2025kolmogorov}.

\subsection{Discontinuity-aware Kolmogorov–Arnold network}
\label{sub:2.3}
Kolmogorov-Arnold representation theorem states that any continuous multivariable function $f(\bm{x})=f(x_1,\cdots,x_n)$ on a bounded domain can be expressed as a finite combination of one-dimensional (1D) continuous functions \cite{kolmogorov1961representation}, formalized as (a 2-layer network):
\begin{equation}
	f(\bm{x}) \approx \sum_{q=1}^{2 n+1} {\phi}_{1,q}^{2}\left(\sum_{p=1}^n \phi_{q,p}^{1}\left(x_p\right)\right),
	\label{eq: kan}
\end{equation}
where $\phi_{q,p}^{1}$ is the learnable univariate continuous activation function at the first layer, with $q$ and $p$ indexing the output and input, respectively, and ${\phi}_{1,q}^{2}$ is the activation function at the second layer, with $q$ indexing its input and producing a single output. This theorem provides a foundation for dimensionality reduction in high-dimensional function approximation. Based on this theorem, a Kolmogorov–Arnold network (KAN) \cite{liu2024kan} is developed as a promising alternative to MLP. An $l$-layer KAN can be expressed as:

\begin{equation}
	\operatorname{KAN}(\bm{x})=\left(\bm{\Phi}_{l} \circ \bm{\Phi}_{l-1} \circ \cdots \circ \bm{\Phi}_1\right) \bm{x},
\end{equation}
where $\bm{\Phi}_{l}=\{\phi_{q,p}^{l}(\bm{x})\}$ represents the function matrix of the $l$-th layer (parameterized by B-splines), with $q$ and $p$ indexing the output and input, respectively, and $\circ$ denotes the composition of functions, indicating that the output of one function is used as the input of another. Studies have demonstrated that much smaller KAN can achieve comparable or better accuracy than larger MLP in PDE solving. Therefore, KAN-based methods can reduce the number of network parameters and improve real-time computational efficiency \cite{liu2024kan}.

\begin{figure}[!h]
	\centering
	\includegraphics{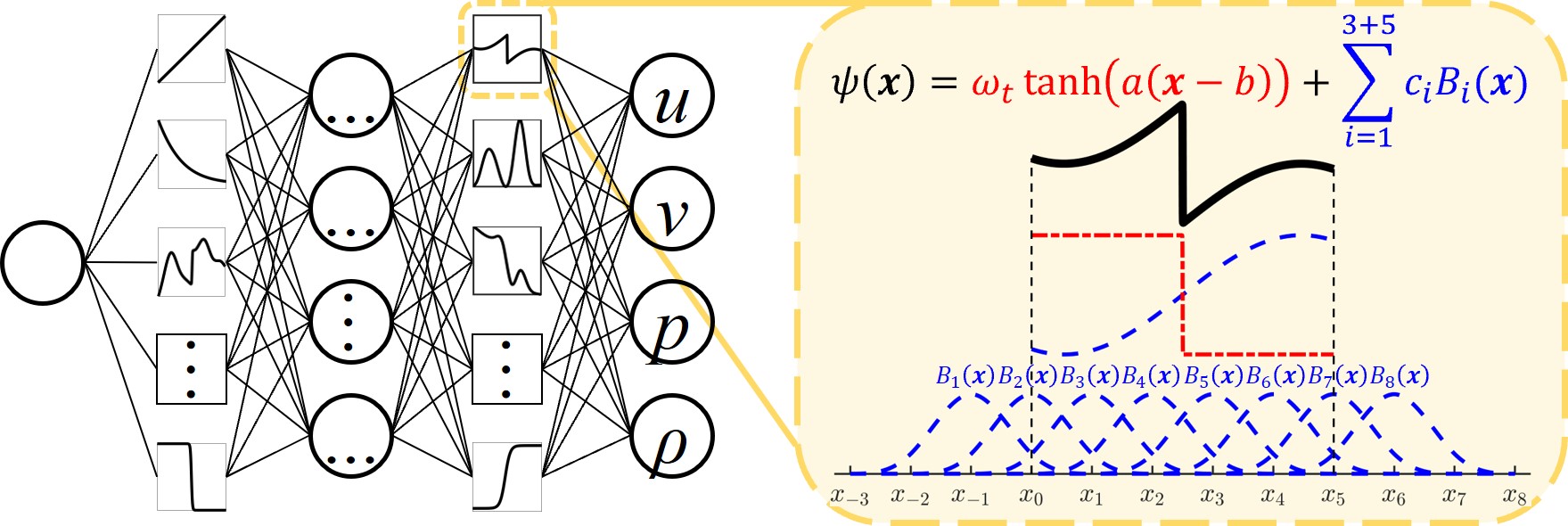}
	\caption{Left: Schematic of the DKAN architecture. Right: The discontinuous-aware activation function consists of parameterized B-splines and DyT functions.}
	\label{fig: DKAN}
\end{figure} 

However, KAN shares the limitations of the conventional MLP in modeling discontinuous features due to its continuity assumption. Although it has been theoretically demonstrated that the Kolmogorov theory exhibits the potential to represent discontinuous functions \cite{ismayilova2024kolmogorov}, practical algorithms to bridge this gap have not yet been developed. DKAN is then designed as a numerical implementation of this theory for automatic discontinuity approximation:

\begin{equation}
	\operatorname{DKAN}(\bm{x})=\left(\bm{\Psi}_{l} \circ \bm{\Psi}_{l-1} \circ \cdots \circ \bm{\Psi}_1\right) \bm{x},
\end{equation}
where $\bm{\Psi}_{l}=\{\psi_{q,p}^{l}(\bm{x})\}$ denotes the learnable discontinuous-aware activation function matrix of the $l$-th layer, with $q$ and $p$ indexing the output and input, respectively. This activation function consists of Dynamic Tanh (DyT) \cite{zhu2025transformers} and parameterized B-spline basis functions \cite{liu2024kan}, enabling accurate approximation of continuous or discontinuities, as shown in Fig. \ref{fig: DKAN}, and is defined as:

\begin{equation}
	\psi (\bm{x})=\operatorname{DyT}(\bm{x})+\phi(\bm{x})=
	\underbrace{{w_t}\tanh(w_a(\bm{x}-w_b))}_{\text{DyT basis}}+
	\underbrace{{w_s}\sum\limits_{i}^{k+g}{w_{c_i}}{S_i}(\bm{x})}_{\text{Spline basis}},
\end{equation}
where $w_a$, $w_b$, $w_{c_i}$, $w_t$ and $w_s$ are trainable parameters, and ${S_i}(\bm{x})$ is the spline basis function defined by the $k$-th order polynomial with $g$ grid points. 

\subsection{Mesh transformation}
\label{sub:2.4}
\begin{figure}[!h]
	\centering
	\includegraphics{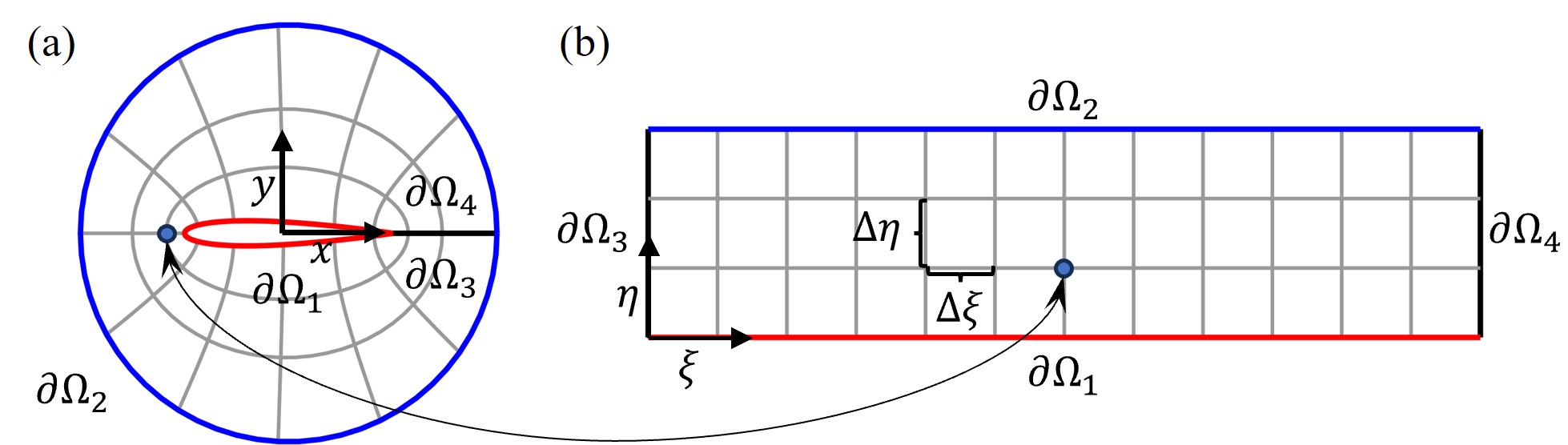}
	\caption{Schematic of the body-fitted coordinate system. (a) Physical space. (b) Computational space.}
	\label{fig: mesh transform}
\end{figure}

Previous studies observed that PINNs suffer from poor convergence and accuracy when applied to problems with sharp spatial transitions or rapid time evolution, such as flow around airfoils \cite{zang2020weak,meng2020ppinn}. To improve convergence for complex geometries, a nonlinear invertible mesh transformation is applied, stretching regions of high mesh resolution and compressing regions of low mesh resolution, as shown in Fig. \ref{fig: mesh transform}. This approach has been successfully used to predict the subsonic steady flow around fully parameterized airfoils \cite{cao2024solver}. $\partial \Omega_1$ and $\partial \Omega_2$ represent the no-slip wall and far-field boundaries, respectively, and $\partial \Omega_3$ and $\partial \Omega_4$ coincide on the physical plane with periodic boundary conditions. The structured body-fitted physical coordinates $(x_{i,j},y_{i,j})$ are transformed into a uniform computational grid $(\xi_{i,j},\eta_{i,j})$ via point-to-point mapping, where $\xi_{i, j}=i \Delta \xi$ and $\eta_{i, j}=j \Delta \eta$, with $\Delta \xi=\Delta \eta=1$ for $i=1, \ldots, N_{\xi}$ and $j=1, \ldots, N_\eta$. Here, $N_\xi$ and $N_\eta$ denote the number of discrete grid points in the $\xi$ and $\eta$ directions, corresponding to the tangential and normal directions relative to the airfoil surface, respectively.

The geometric metrics are computed via finite differences between the body-fitted domain and the uniform one. For interior grid points, second-order central differences are employed:
\begin{equation}
	\begin{aligned}
		&x_{\xi}=\frac{x_{i+1, j}-x_{i-1, j}}{2 \Delta \xi}, \quad y_{\xi}=\frac{y_{i+1, j}-y_{i-1, j}}{2 \Delta \xi},\\
		&x_\eta=\frac{x_{i, j+1}-x_{i, j-1}}{2 \Delta \eta}, \quad y_\eta=\frac{y_{i, j+1}-y_{i, j-1}}{2 \Delta \eta},
	\end{aligned}
	\label{center}
\end{equation}
where the subscripts indicate partial derivatives. At the boundaries, first-order differences are applied. The corresponding first-order inverse metrics are given by:
\begin{equation}
	\xi_x = \frac{y_\eta}{J}, \quad \eta_x = -\frac{y_\xi}{J}, \quad	\xi_y = -\frac{x_\eta}{J},  \quad \eta_y = \frac{x_\xi}{J},
	\label{Jacobian}
\end{equation}
where $J = \frac{\partial(x,y)}{\partial(\xi,\eta)}=x_\xi y_\eta - x_\eta y_\xi$ is the Jacobian determinant. Based on the chain rule of differential calculus, the second-order inverse metrics are computed by:
\begin{equation}
	\begin{aligned}
		\xi_{xx} &= \xi_x (\xi_x)_\xi + \eta_x (\xi_x)_\eta, & \eta_{xx} &= \xi_x (\eta_x)_\xi + \eta_x (\eta_x)_\eta, \\
		\xi_{xy} &= \xi_y (\xi_x)_\xi + \eta_y (\xi_x)_\eta, & \eta_{xy} &= \xi_y (\eta_x)_\xi + \eta_y (\eta_x)_\eta,\\
		\xi_{yy} &= \xi_y (\xi_y)_\xi + \eta_y (\xi_y)_\eta, & \eta_{yy} &= \xi_y (\eta_y)_\xi + \eta_y (\eta_y)_\eta, 
	\end{aligned}
\end{equation}
where the derivatives of the inverse metrics, such as $(\xi_x)_\xi$, are computed by applying the difference operator defined in Eq. \eqref{center} to the terms in Eq. \eqref{Jacobian}. Finally, the first- and second-order derivatives in the physical space are computed from the uniform computational domain to formulate the PDE constraints:

\begin{equation}\label{eq:chain-matrix}
	\begin{pmatrix}
		\partial_x\\
		\partial_y\\
		\partial_{xx}\\
		\partial_{xy}\\
		\partial_{yy}
	\end{pmatrix}
	=
	\begin{pmatrix}
		\xi_x & \eta_x & 0 & 0 & 0\\
		\xi_y & \eta_y & 0 & 0 & 0\\
		\xi_{xx} & \eta_{xx} & \xi_x^2 & 2\,\xi_x\eta_x & \eta_x^2 \\
		\xi_{xy} & \eta_{xy} & \xi_x\xi_y & \xi_x\eta_y+\eta_x\xi_y & \eta_x\eta_y \\
		\xi_{yy} & \eta_{yy} & \xi_y^2 & 2\,\xi_y\eta_y &  \eta_y^2
		
	\end{pmatrix}
	\begin{pmatrix}
		\partial_\xi\\
		\partial_\eta\\
		\partial_{\xi\xi}\\
		\partial_{\xi\eta}\\
		\partial_{\eta\eta}
	\end{pmatrix}.
\end{equation}

\subsection{Learnable local artificial viscosity}
\label{sub:2.5}
Artificial viscosity (AV) is a well-established approach in computational fluid dynamics for shock capturing, evolving in various forms to improve numerical stability and accuracy \cite{white1973new,caramana1998formulations}. It has also been introduced into PINNs as a physics modification strategy, incorporating an additional diffusion term to smooth sharp spatial transitions and rapid temporal evolution, albeit at the expense of numerical accuracy \cite{wassing2025adopting,wassing2024physics,fuks2020limitations,coutinho2023physics}. The artificial-viscosity form of the compressible Euler equations can be formulated as:

\begin{equation}
	\begin{gathered}
		\frac{\partial \bm{Q}}{\partial t}+\nabla \cdot \bm{F} =\mu \nabla^2 \bm{Q} \\ 
		\bm{Q}=\left(\begin{array}{c}\rho \\ \rho \bm{u} \\ E\end{array}\right), \bm{F}=\left(\begin{array}{c}\rho \bm{u} \\ \rho \bm{u} \otimes \bm{u} + p\bm{I} \\ (E+p) \bm{u} \end{array}\right)
	\end{gathered}
	\label{eq: flux}
\end{equation}
where $\bm{Q}$ is the conservative state vector, $\bm{F}$ represents the flux vector, $\rho$ is the density, $\bm{u}$ is the velocity vector, $E=\frac{p}{\gamma-1}+\frac{1}{2} \rho|\bm{u}|^2$ is the total energy with $p$ as the pressure and $\gamma=1.4$ as the ratio of specific heats, and $\mu \ge 0$ denotes the artificial viscosity coefficient controlling the magnitude of dissipation.

However, selecting an appropriate value for $\mu$ is challenging: an excessively large value induces over-dissipation and reduces accuracy, whereas an insufficient value fails to mitigate numerical instability and resolve discontinuous solutions \cite{coutinho2023physics}. In numerical methods, $\mu$ is typically chosen to be proportional to the grid spacing $\Delta \bm{x}$ and the magnitude of the gradient of the solution \cite{margolin2023artificial}. Theoretically, a sufficiently fine mesh allows numerical schemes to resolve discontinuities fully, but the computational resources required in practice are prohibitive.

For PINNs, $\mu$ is also typically determined empirically, and its dependence on system properties, such as the training set and network size, has not been explored \cite{abbasi2025challenges}. Although architecture modification methods have demonstrated that increasing network complexity can resolve problems exhibiting discontinuities, their applicability remains limited to 1D cases due to the large computational resource requirements \cite{rodriguez2022physics,anagnostopoulos2024residual,jagtap2020extended,wang2021eigenvector,bruno2022fc,mojgani2023kolmogorov,jagtap2020adaptive}. Consequently, artificial viscosity remains an essential tool for balancing accuracy and efficiency. Besides being specified, the viscosity coefficient $\mu$ can also be embedded within the network to adapt to varying solution complexities \cite{coutinho2023physics} and applied globally \cite{raissi2019physics} or locally \cite{wassing2025adopting,wassing2024physics} in regions with sharp transitions. However, accurately capturing these regions across different scenarios remains challenging, and this approach is currently limited to simple cases.

In this study, a learnable local AV method is developed to enhance the modeling of shock-wave properties. The viscosity coefficient $\mu$ is divided into two parts: magnitude and region of application. Its magnitude is scaled by the spectral radius $r(\bm{x})$ of the flux Jacobians in Eq. \eqref{eq: flux}, following classical scalar dissipation schemes such as the Jameson-Schmidt-Turkel scheme \cite{jameson1981numerical}. The region of application is limited to the areas identified by a sensor function $s(\bm{x})$. Therefore, $\mu$ is defined as:
\begin{equation}
	\mu=w_\nu r(\bm{x}) s(\bm{x}),
\end{equation}
where $w_\nu$ is treated as a single non-negative parameter learned during training. Finally, the PDE loss in Eq. \eqref{eq: loss} is redefined as:
\begin{equation}
	\mathcal{L}_{pde}=\left\| \left( \frac{\partial \bm{Q}}{\partial t}+\nabla \cdot \bm{F}-\mu \nabla^2 \bm{Q} \right) V \right\|_2^2 / \left\|V \right\|_2^2,
\end{equation}
where $V$ is the grid volume in the physical coordinate system, weighting the residuals to avoid bias toward regions with dense mesh refinement \cite{song2025vw}.
\section{Results}
\label{sub:3}
This section evaluates the performance of the proposed method in solving PDEs with sharp spatial transitions or fast temporal evolution across various types of equations and dimensions, and benchmarks its accuracy and the number of parameters compared to state-of-the-art techniques. As a preliminary test, we first present results obtained for the standard 1D inviscid Burgers' equation. Then, the method is applied to transonic flow around an airfoil with normal shocks on the airfoil surface. Finally, it is evaluated for supersonic airfoil flow, which involves additional features such as upstream detached bow shocks and trailing-edge oblique shocks.

Due to sharp transitions, three error metrics are employed to measure the accuracy of the solution, defined as: 
\begin{equation}
	\begin{aligned}
		R_1=\frac{\left\| u_{true}-u \right\|_1}{\left\| u_{true} \right\|_1}, \quad R_2=\frac{\left\| u_{true}-u \right\|_2}{\left\| u_{true} \right\|_2}, \quad R_{front}=\frac{\left\| x_{front,true}-x_{front} \right\|_2}{\left\| x_{front,true} \right\|_2},
	\end{aligned}
	\label{eq: three metrics}
\end{equation}
where $\left\|\cdot \right\|_1$ represents the $L_1$ norm, $R_{front}$ marks the relative position error of the shock front between the exact and the PINN solutions, and $x_{front}=\underset{x}{\operatorname{argmin}} \frac{\partial u}{\partial x}$.
\subsection{Inviscid Burgers' equation}
\label{sub:3.1}
The 1D inviscid Burgers' equation with sinusoidal initial and homogeneous boundary conditions is a classical benchmark of PINNs, commonly used to describe shock-waves, given by:
\begin{equation}
	\begin{aligned}
		\frac{\partial u}{\partial t}+\frac{\partial (u^2/2)}{\partial x}=0, \quad &x\in [0,2], \quad t\in [0,1]\\
		u(x,0)=-\sin( \pi(x-1)) ,\quad &x\in [0,2],\\
		u(0,t)=u(2,t)=0, \quad &t\in [0,1].
	\end{aligned}
	\label{eq: burgers_equ}
\end{equation}
Here, the computational domain is uniformly discretized with 200 points in both space and time. The sensor and the spectral radius of the flux Jacobians in Eq. \eqref{eq: burgers_equ} are defined as:

\begin{equation}
	\begin{aligned}
		s(\bm{x})&=\operatorname{tanh}(\max (0, k_{\text {shock }, 1} \left[-{u}/{|u|}-k_{\text {shock},0}\right] )),\\
		r(\bm{x})&=|u|,\\
	\end{aligned}
\end{equation}
where $k_{\text {shock},0}=0.015$ and $k_{\text {shock }, 1}=1$. The mesh transformation becomes trivial, as the transformed and computational domains are identical besides linear scaling and translation. 

The DPINN method consists of a KAF with 64 embedding frequencies and an output dimension of 5, followed by 2 DKAN layers (5 neurons/layer). The other methods employ a 64-frequency RFF followed by 4 weight-normalized fully-connected layers (30 neurons/layer) or 2 KAN layers (5 neurons/layer) for MLP- and KAN-based models, respectively, each comprising standard, global AV, and local AV variants ($\mu = 3 \times 10^{-3}$). These networks are trained using the AdamW optimizer for 50000 epochs with a learning rate of 0.001, and the loss term weights are specified as $\lambda_{pde}=\lambda_{ic}=\lambda_{bc}=1$. Detailed training hyperparameters are provided in Appendix A. Adaptive weighting algorithms can automatically adjust loss weights during training \cite{xiang2022self}. However, for the problems considered herein, no significant improvement was observed by using such algorithms compared to constant values.

\begin{figure}[!h]
	\centering
	\includegraphics[width=0.6\linewidth]{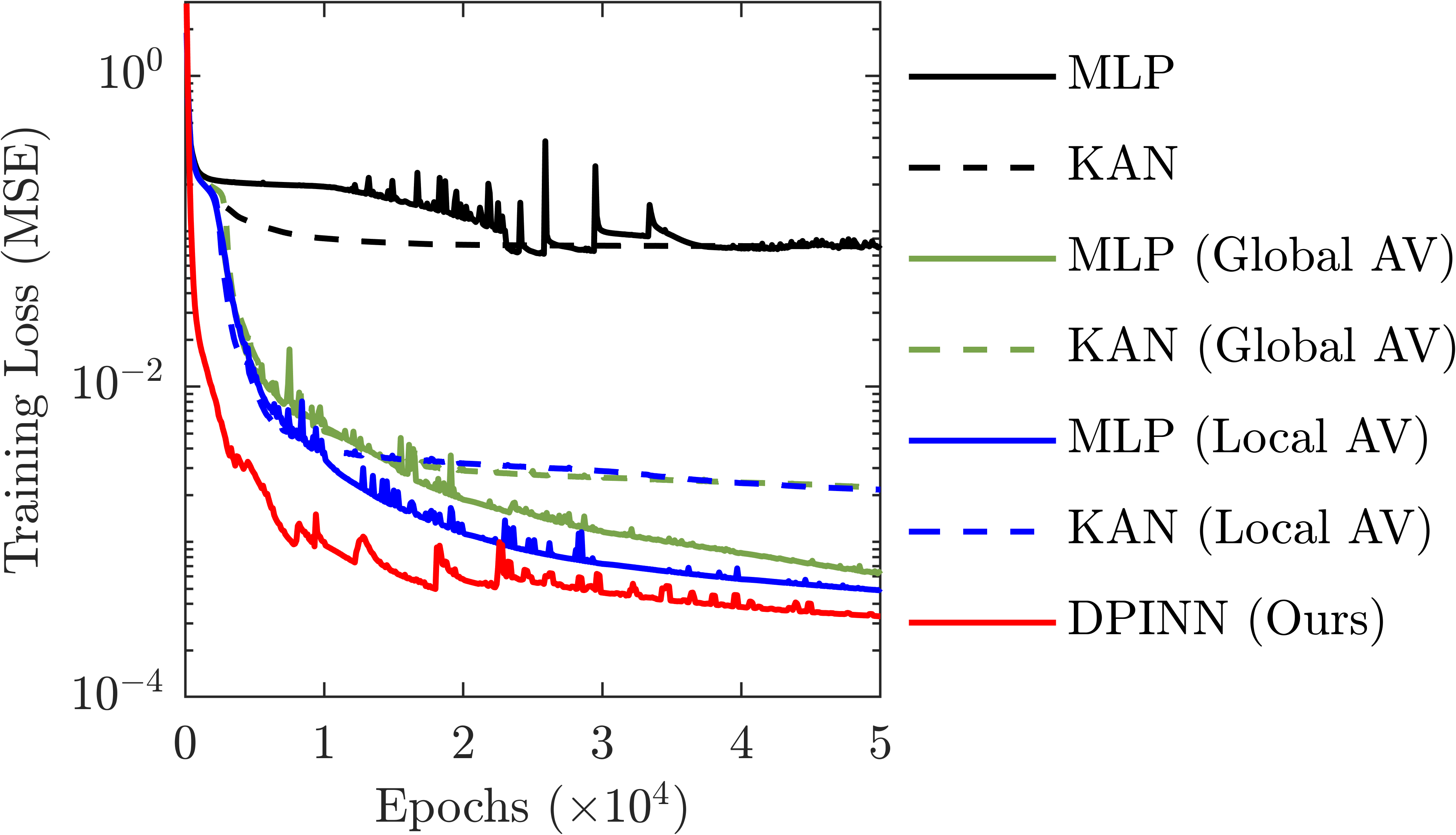}
	\caption{Evolution of the loss function with training epochs.}
	\label{fig: burger_loss}
\end{figure}

\begin{figure}[!h]
	\centering
	\includegraphics{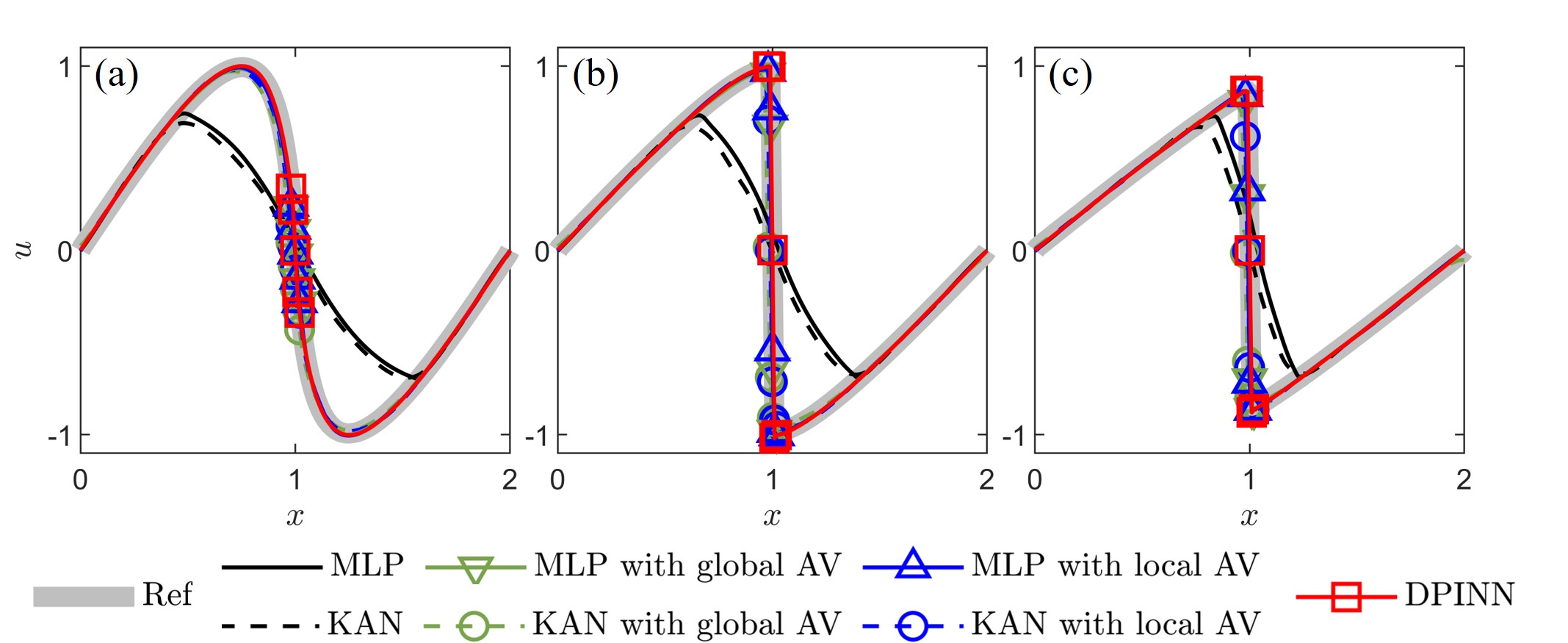}
	\caption{The solutions at different time steps for various methods, with markers placed at $x=1$ and its four adjacent points. (a) $t=0.25$; (b) $t=0.5$; (c) $t=0.75$.}
	\label{fig: burger_solution}
\end{figure}

Fig. \ref{fig: burger_loss} displays the training loss histories for the above methods. Conventional MLP- and KAN-based methods fail to converge adequately due to discontinuities \cite{liu2024discontinuity}. In contrast, DPINN, MLP with AV, and KAN with AV reduce the loss by over two orders of magnitude, with DPINN exhibiting the fastest convergence and the lowest final loss. Solutions at different times are shown in Fig. \ref{fig: burger_solution}, where the reference solutions are obtained using the WENO method. While MLP and KAN methods fail to capture discontinuities, and their variants with global or local AV produce smooth low-resolution shock fronts due to viscous dissipation, DPINN accurately resolves discontinuities with errors below $1\%$ across all metrics, requiring less than $1/5$ the parameters of MLP-based methods (Table \ref{T: burgers_acc}).

In terms of the trade-off between accuracy and efficiency, DPINN requires $4.69 \times$ and $2.96 \times$ the training wall-clock time of MLP and MLP with local AV (best baseline), respectively. Nevertheless, it yields $50.22 \times$ and $4.05 \times$ higher $L_2$ accuracy while maintaining a comparable inference time. Furthermore, DPINN outperforms KAN, KAN with global AV, and KAN with local AV, achieving significant accuracy improvements without introducing significant computational overhead.

\begin{table}[!h]
	\centering
	\renewcommand\arraystretch{1.1}
	\resizebox{0.9\textwidth}{!}{
		\begin{tabular}{@{}c | c | ccc | c | cc@{}}
			\hline
			\multirow{4}{*}[1.25ex]{Model} & 
			\multirow{4}{*}[1.25ex]{Final Loss} & 
			\multicolumn{3}{c|}{\multirow{2}{*}[0.5ex]{Relative Error (\%)}} & 
			\multirow{4}{*}[1.25ex]{Params} & 
			\multicolumn{2}{c}{\multirow{2}{*}[0.5ex]{Wall-clock Time (s)}} \\[-0.8ex] 
			
			& & \multicolumn{3}{c|}{} & & \multicolumn{2}{c}{} \\[-0.8ex] 
			
			\cline{3-5} \cline{7-8}
			
			&  & \multirow{2}{*}[0.5ex]{$R_1$} & \multirow{2}{*}[0.5ex]{$R_2$} & \multirow{2}{*}[0.5ex]{$R_{front}$} & & Training & Inference \\[-0.8ex] 
			
			& & & & & & ($\times 10^2$) & ($\times 10^{-3}$) \\[-0.8ex] 
			\hline
			
			MLP 			& $7.72 \times 10^{-2}$ & 26.55 & 44.19 & 2.11 & 4835 & 6.89 & 3.10\\
			KAN 			& $8.02 \times 10^{-2}$ & 27.13 & 43.61 & 7.21 & 689 & 18.06 & 4.33\\
			MLP (Global AV)	& $6.21 \times 10^{-4}$ & 1.55 & 4.79 & 1.16 & 4835 & 10.78 & 3.24\\
			KAN (Global AV)	& $2.24 \times 10^{-3}$ & 3.04 & 7.57 & 1.72 & 689 & 30.98 & 4.67\\
			MLP (Local AV)	& $5.02 \times 10^{-4}$ & 1.29 & 3.56 & 0.94 & 4835 & 10.93 & 3.26\\
			KAN (Local AV) 	& $2.18 \times 10^{-3}$ & 2.57 & 5.84 & 1.48 & 689 & 31.38 & 4.74\\
			\hline
			DKAN			& $4.47 \times 10^{-4}$ & 0.82 & 0.94 & 0.45 & 491 & 23.20 & 6.84\\
			DKAN+Fourier	& $4.00 \times 10^{-4}$ & 0.82 & 0.91 & 0.43 & 770 & 19.61 & 4.68\\
			DKAN (Local AV)	& $5.14 \times 10^{-4}$ & 0.96 & 1.31 & 0.65 & 492 & 36.18 & 5.80\\
			\rowcolor{gray!20} DPINN (Ours) & $3.32 \times 10^{-4}$ & 0.80 & 0.88 & 0.38 & 771 & 32.34 & 5.06\\
			\hline
		\end{tabular}
	}
	\caption{Inviscid Burgers' equation results, measured on an Nvidia GeForce RTX-4090 with 24 GB of memory.}
	\label{T: burgers_acc}
\end{table}

\begin{figure}[!h]
	\centering
	\includegraphics[width=0.7\linewidth]{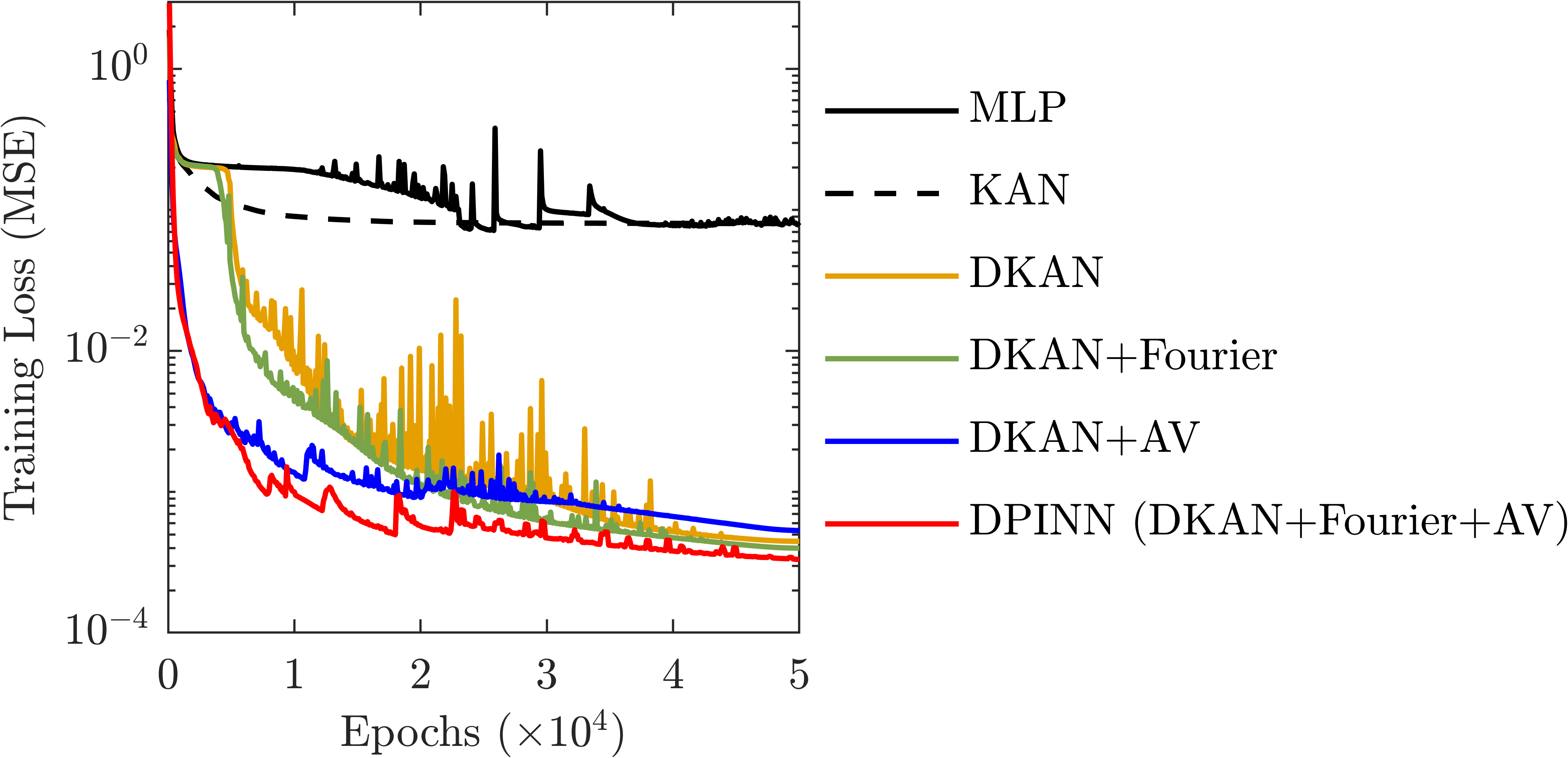}
	\caption{Training loss evolution in the ablation study for the 1D Burgers' equation.}
	\label{fig: Ablation_burgers}
\end{figure}

Furthermore, a detailed ablation study is performed to evaluate the contributions of individual components, with loss histories and results summarized in Fig. \ref{fig: Ablation_burgers} and Table \ref{T: burgers_acc}, respectively. As an architectural modification strategy, DKAN is capable of achieving a substantially lower loss than traditional models and accurately approximating discontinuities without auxiliary support, despite significant training oscillations caused by shock-capturing instabilities. Fourier-feature embedding mitigates spectral bias to resolve high-frequency discontinuities, stabilizing the optimization process. While artificial viscosity significantly accelerates and stabilizes convergence, its failure to decay to a negligible magnitude hinders the accurate resolution of shocks. The proposed DPINN incorporates a Fourier-feature embedding layer to mitigate spectral bias, DKAN to model shock-wave properties, and local artificial viscosity to stabilize the algorithm near discontinuities, enabling accurate shock capturing.

\subsection{One-dimensional Riemann problem}
\label{sub:3.15}
The Sod shock tube is a classic 1D Riemann problem governed by the Euler equations. An initial density and pressure discontinuity at the center of the computational domain evolves into a shock wave propagating into the low-pressure region, an expansion wave traveling into the high-pressure region, and a contact discontinuity in between. The computational domain $(x,t) \in [0,1] \times [0,0.2]$ is uniformly discretized with 200 points in each direction. The initial conditions are given by:
\begin{equation}
	(\rho, u, p)|_{t=0} = 
	\begin{cases} 
		(1.0,0.0,1.0), & x \in [0, 0.5), \\ 
		(0.125,0.0,0.1), & x \in [0.5, 1.0].
	\end{cases}
\end{equation}

Compared to the Sod case, the Lax shock tube is a more challenging Riemann problem due to its larger initial velocity, which produces stronger shocks and a faster contact discontinuity. The computational domain $(x,t) \in [0,1] \times [0,0.14]$ is uniformly discretized with 200 spatial and 140 temporal points, respectively, with initial conditions given by:
\begin{equation}
	(\rho, u, p)|_{t=0} = 
	\begin{cases} 
		(0.445, 0.689, 3.528), & x \in [0, 0.5), \\ 
		(0.5, 0.0, 0.571), & x \in [0.5, 1.0].
	\end{cases}
\end{equation}

For both 1D Riemann problems, DPINN consists of a KAF with 128 embedding frequencies and an output dimension of 3, followed by 2 DKAN layers (12 neurons/layer). The other methods, comprising standard, global AV, and local AV variants ($\mu = 4 \times 10^{-4}$), employ a 128-frequency RFF layer followed by five weight-normalized fully-connected layers (96 neurons per layer). These networks are trained using the AdamW optimizer for 20000 epochs at a learning rate of 0.001, with loss weights set to $\lambda_{pde}=\lambda_{ic}=\lambda_{bc}=1$.

\begin{figure}[!h]
	\centering
	\includegraphics[width=\linewidth]{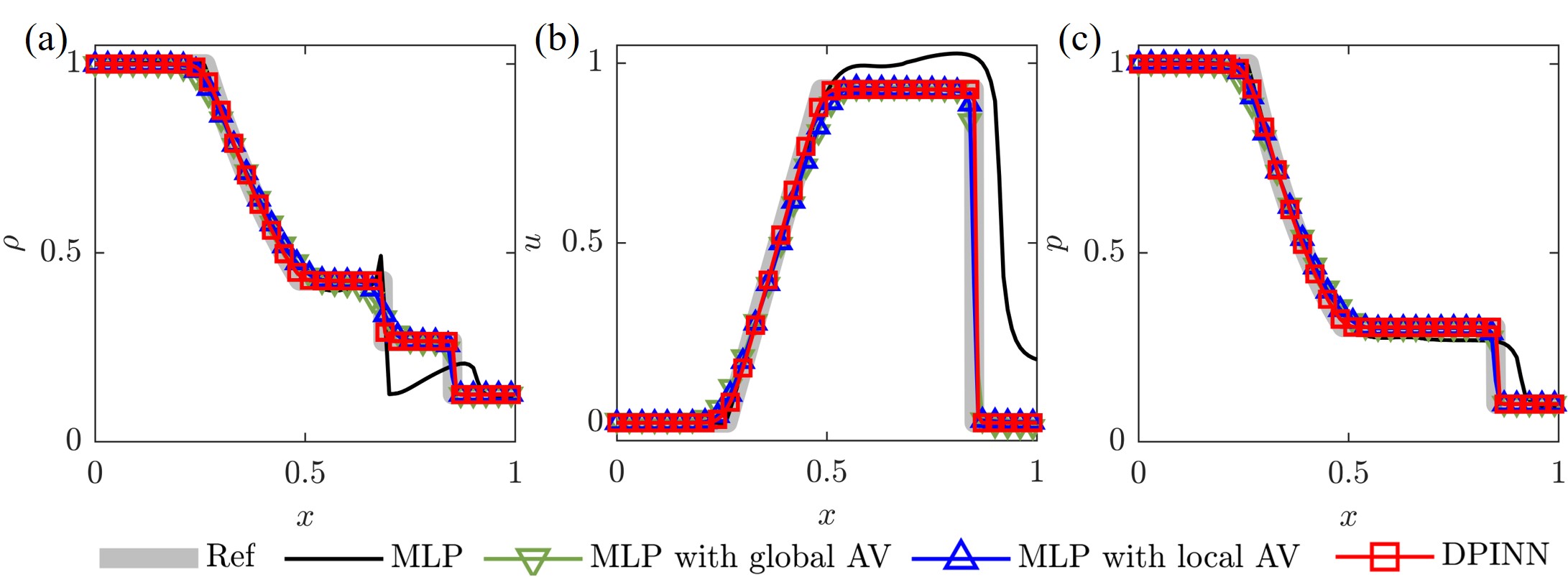}
	\caption{Sod Shock Tube: solution at $t$ = 0.2.}
	\label{fig: SST_line}
\end{figure}

\begin{figure}[!h]
	\centering
	\includegraphics[width=\linewidth]{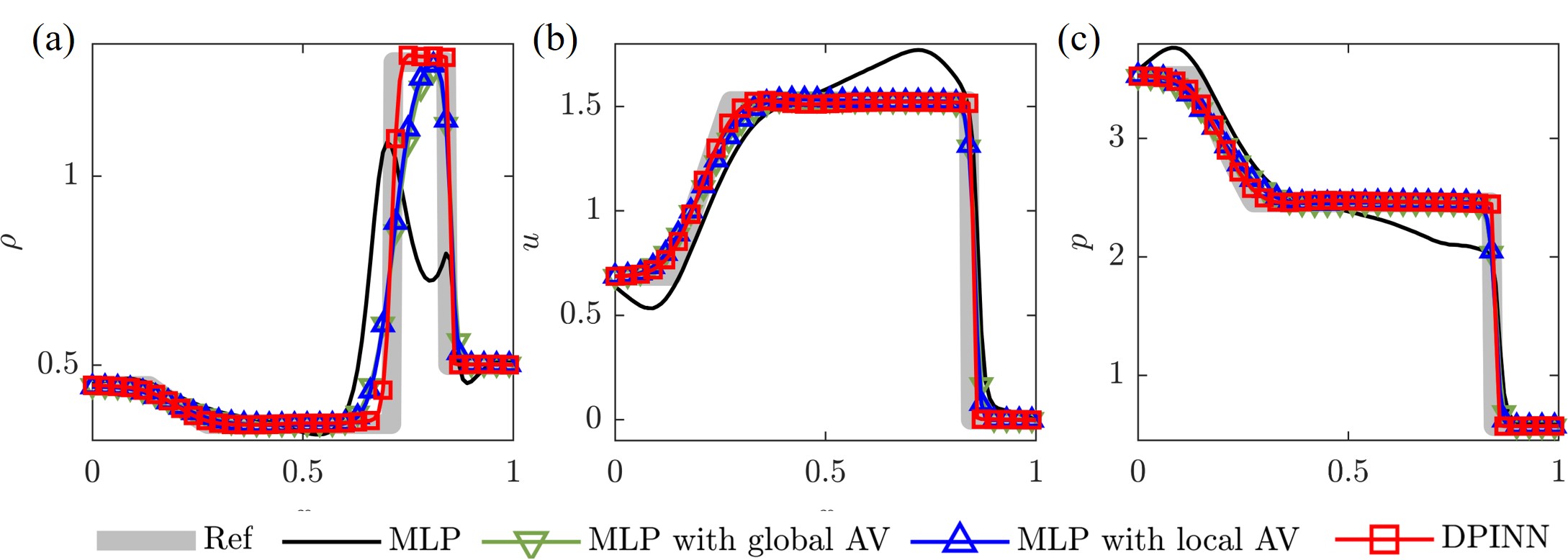}
	\caption{Lax Shock Tube: solution at $t$ = 0.14.}
	\label{fig: Lax_line}
\end{figure}

Fig. \ref{fig: SST_line} and Fig. \ref{fig: Lax_line} show quantitative comparisons between the exact solutions and the PINN predictions for the Sod and Lax cases, respectively. Conventional MLP-based methods exhibit significant deviations near discontinuities. While MLPs with AV capture smooth shock profiles in the Sod case, they fail to resolve the Lax case, resulting in an excessively dissipative flow field. In contrast, DPINN demonstrates consistent superiority in both scenarios, achieving accurate shock capturing. The predicted flow fields over the entire computational domain are provided in Appendix C (see Fig. \ref{fig: SST_contourf} and Fig. \ref{fig: Lax_contourf}).

\subsection{Transonic flows around an airfoil}
\label{sub:3.2}
The proposed method is then demonstrated on a 2D transonic NACA0012 airfoil flow. In this scenario, the flow locally accelerates to supersonic speeds, forming normal shock-waves on the airfoil surface and resulting in shock-boundary layer interactions that significantly influence lift, drag, and overall aerodynamic performances. Accurate and efficient simulation of these phenomena is essential for the aerodynamic optimization of transonic aircraft. The governing equations are the 2D compressible Euler equations. The Mach number is defined as $Ma=|\bm{u}|/c$, with the speed of sound given by $c=\sqrt{\gamma p/\rho}$. The problem is solved within a finite circular domain of diameter 30 and chord length 1, as shown in Fig. \ref{fig: mesh transform}, with $N_{\xi} \times N_{\eta}=200 \times 100$ points. The high-fidelity reference solution is obtained using a second-order finite volume method on a fine mesh $N_{\xi} \times N_{\eta}=800 \times 400$. 

The DPINN method consists of a KAF with 128 embedding frequencies and an output dimension of 4, followed by 3 DKAN layers (12 neurons/layer). The other methods employ a 128-frequency RFF followed by 4 weight-normalized fully-connected layers (128 neurons/layer) or 3 KAN layers (12 neurons/layer) for MLP- and KAN-based models, respectively, each comprising standard, global AV, and local AV variants ($\mu = 1.25 \times 10^{-3}$). These networks are trained using the SOAP optimizer for 50000 epochs with a learning rate of 0.003, and the loss term weights are set to $\lambda_{pde}= 2 \times 10^4$ and $\lambda_{bc}=1$, without considering the initial condition loss.

In this case, the sensor and the spectral radius of the flux Jacobians in Eq. \eqref{eq: flux} are constructed as:
\begin{equation}
	\begin{aligned}
		s(x,y)&=s_{\text {shock}}(x,y)s_{\text{stag}}(x,y),\\
		r(x,y)&=|\bm{u}|+c,\\
	\end{aligned}
	\label{eq:qb}
\end{equation}
where $s_{\text {shock}}(x,y)$ is the shock sensor and $s_{\text{stag}}(x,y)$ denotes the stagnation-point sensor. For $s_{\text{shock}}(x,y)$, as pressure generally increases along the flow direction across shock waves, the dot product of the normalized velocity ${\bm{u}}/|\bm{u}|$ and the pressure gradient $\bm{\nabla} p$ is positive near shock areas. To distinguish shocks from steep stagnation-point pressure gradients (e.g., at an airfoil leading edge), $s_{\text{stag}}(x,y)$ is introduced, as momentum gradients $\bm{\nabla}\rho u$ and $\bm{\nabla}\rho v$ are significant near stagnation points but remain small in shock regions. They are defined as:

\begin{equation}
	\begin{aligned}
		s_{\text {shock}}(x,y)&=\operatorname{tanh} (\max (0, k_{\text {shock }, 1} \left[ ({\bm{u}}/{|\bm{u}|}) \cdot \bm{\nabla} p-k_{\text {shock }, 0}\right] )),\\
		s_{\text{stag}}(x,y)&=\operatorname{Sigmoid}\left(k_{\text{stag}, 1}\left[ k_{\text{stag}, 0}-(|\bm{\nabla} \rho u|+|\bm{\nabla} \rho v|)\right] \right),
	\end{aligned}
\end{equation}
where $\operatorname{Sigmoid}(x)=1/[1+\exp(-x)]$, and the hyperparameters $k_{shock,0}=0.5$, $k_{shock,1}=5$, $k_{stag,0}=3$, and $k_{stag,1}=4$ control the threshold and steepness of $s_{\text {shock}}$ and $s_{\text {stag}}$, respectively. A detailed parameter sensitivity analysis is provided in Appendix B. For details on the sensors, the interested reader is referred to \cite{wassing2024physics}.

The flow around the NACA0012 airfoil is analyzed under transonic conditions ($Ma_{\infty}=0.7$) at angles of attack $\alpha=0^{\circ}$ and $\alpha=4^{\circ}$ for scenarios with and without shocks, respectively. The far-field conditions are specified as:
\begin{equation}
	u_{\infty}=\sqrt{\gamma} Ma_{\infty} \cos(\alpha), \quad v_{\infty}=\sqrt{\gamma} Ma_{\infty} \sin(\alpha), \quad p_{\infty}=1, \quad \rho_{\infty}=1,
	\label{condition}
\end{equation}

\begin{figure}[!h]
	\centering
	\includegraphics[width=0.8\linewidth]{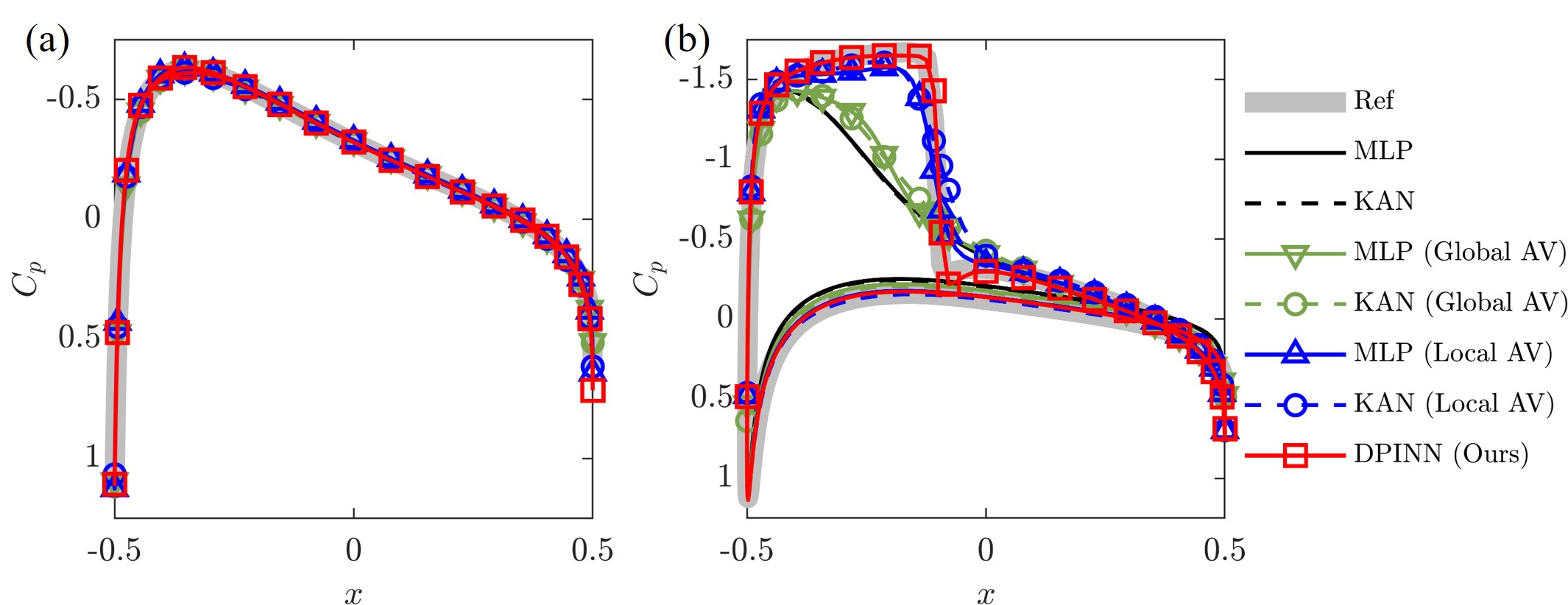}
	\caption{The pressure coefficient distributions around the NACA0012 airfoil at $Ma=0.7$ and $\alpha=0^{\circ}$ and $\alpha=4^{\circ}$ for (a) and (b), respectively.}
	\label{fig: transonic cp}
\end{figure}

\begin{table}[htbp]
	\centering
	\renewcommand\arraystretch{1.1}
	\resizebox{\textwidth}{!}{
		\begin{tabular}{@{}c | cc ccc cc | c | cc@{}}
			\hline
			\multirow{3}{*}{Model} &
			\multicolumn{2}{c}{Transonic} &
			\multicolumn{3}{c}{Transonic} &
			\multicolumn{2}{c|}{Supersonic} &
			\multirow{3}{*}{Params} &
			\multicolumn{2}{c}{Average Time (s)} \\[-0.4ex] 
			
			\cline{10-11} 
			
			& \multicolumn{2}{c}{($\alpha=0$)}
			& \multicolumn{3}{c}{($\alpha=4$)}
			& \multicolumn{2}{c|}{($\alpha=0$)}
			&
			& Training & Inference \\[-0.4ex] 
			
			\cline{2-8} 
			
			& $R_1$ & $R_2$ & $R_1$ & $R_2$ & $R_{front}$ & $R_1$ & $R_2$ &
			& ($\times 10^3$) & ($\times 10^{0}$) \\[-0.4ex] 
			\hline
			
			MLP
			& 2.01 & 2.65 & 30.18 & 32.86 & 19.46 & 11.04 & 14.71
			& 67208 & 2.65 & 1.15 \\
			KAN
			& 2.83 & 4.47 & 30.06 & 33.08 & 19.07 & 10.82 & 12.56
			& 5036 & 10.90 & 3.65 \\
			MLP (Global AV)
			& 2.31 & 2.76 & 22.29 & 27.88 & 19.46 & 7.70 & 15.55
			& 67208 & 8.11 & 2.90 \\
			KAN (Global AV)
			& 2.59 & 3.96 & 21.37 & 27.09 & 19.07 & 8.64 & 17.18
			& 5036 & 31.74 & 12.10 \\
			MLP (Local AV)
			& 1.58 & 2.45 & 9.45 & 10.73 & 2.31 & 5.38 & 6.17
			& 67208 & 8.40 & 2.93 \\
			KAN (Local AV)
			& 3.21 & 3.91 & 10.45 & 11.57 & 6.46 & 6.84 & 8.22
			& 5036 & 32.24 & 11.51 \\
			\hline
			
			DKAN 
			& 1.92 & 2.87 & 28.07 & 30.09 & 19.46 & 10.76 & 12.44 
			& 4360 & 13.09 & 4.67 \\
			DKAN+Fourier 
			& 1.57 & 2.41 & 28.60 & 30.81 & 19.46 & 10.59 & 11.93 
			& 5784 & 12.02 & 3.94 \\
			DKAN (Local AV) 
			& 2.25 & 2.67 & 4.63 & 5.43 & 1.38 & 5.89 & 6.78
			& 4361 & 40.48 & 13.21 \\
			\rowcolor{gray!20} DKAN (Ours)
			& 1.47 & 2.30 & 3.17 & 3.80 & 0.39 & 4.01 & 4.63
			& 5785 & 34.49 & 10.11 \\
			\hline
		\end{tabular}
	}
	\caption{Results for the transonic and supersonic flows around an airfoil, measured on an Nvidia GeForce RTX-4090 with 24 GB of memory.}
	\label{T: Ma07 acc}
\end{table}

Fig. \ref{fig: transonic cp} and Table \ref{T: Ma07 acc} compare the surface pressure coefficient distributions along the airfoil with and without shock-waves, defined as:

\begin{equation}
	C_p=\frac{p-p_{\infty}}{\frac{1}{2} \rho_{\infty} \left\| \bm{u}_{\infty}\right\|^2_2 },
\end{equation}
where $\bm{u}_{\infty}=[u_{\infty},v_{\infty}]^\top$ is the freestream velocity vector. In the shock-free scenario (Fig. \ref{fig: transonic cp}a), all methods accurately approximate the solution for the complex geometry with $R_1$ and $R_2$ errors below 5\% ($R_{front}$ is not considered). However, in the shock-wave scenario (Fig. \ref{fig: transonic cp}b), conventional MLP and KAN, as well as MLP and KAN with global AV, fail to capture shock structures. While MLP and KAN with local AV produce smooth shock profiles, their resolution remains insufficient for engineering accuracy. In contrast, the proposed DPINN accurately captures shocks with errors below 4\% across all three metrics, significantly outperforming the other methods. 

For model complexity, DPINN requires an order of magnitude fewer parameters than MLP-based models. In terms of the trade-off between accuracy and efficiency, DPINN achieves $8.65\times$ higher $L_2$ accuracy than MLP, at the expense of $13.02\times$ and $8.79\times$ the training and inference times, respectively. Compared with MLP with local AV (the best baseline), DPINN provides $2.82\times$ higher accuracy at the expense of $4.11\times$ the training and $3.45\times$ the inference overhead. Furthermore, DPINN outperforms KAN, KAN with global AV, and KAN with local AV, achieving significant accuracy improvements while maintaining comparable computational overhead.

\begin{figure}[H]
	\centering
	\includegraphics[width=\textwidth]{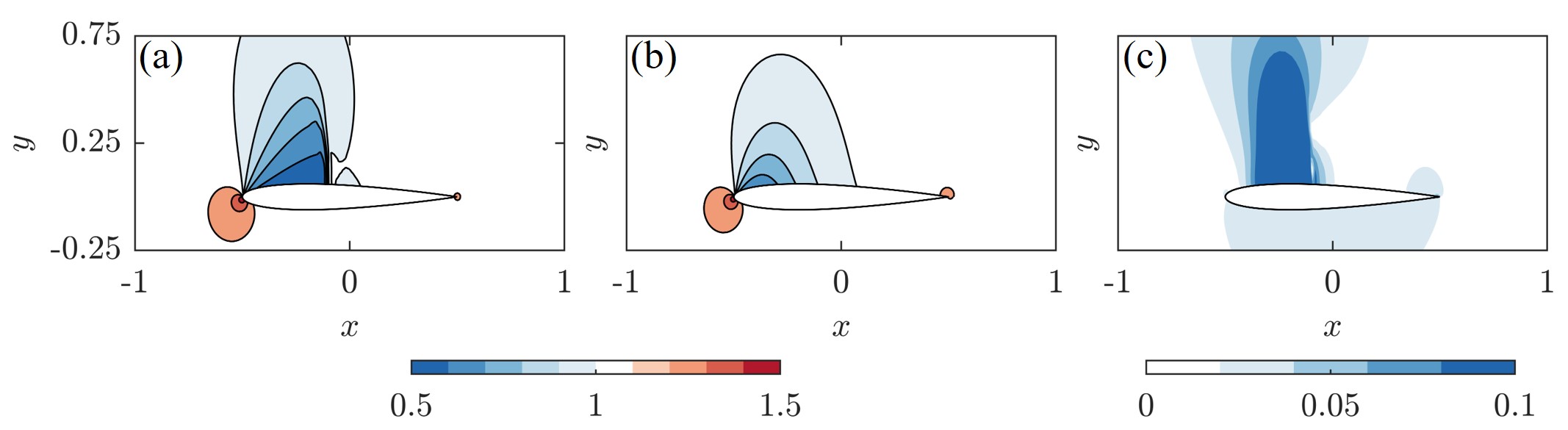}
	\caption{Global pressure field for the transonic flow around the NACA0012 airfoil at $\alpha=4^{\circ}$. (a) High-fidelity solution; (b) Conventional MLP-based method solution; (c) Absolute error.}
	\label{fig: transonic contourf0}
\end{figure}
\begin{figure}[H]
	\centering
	\includegraphics[width=\textwidth]{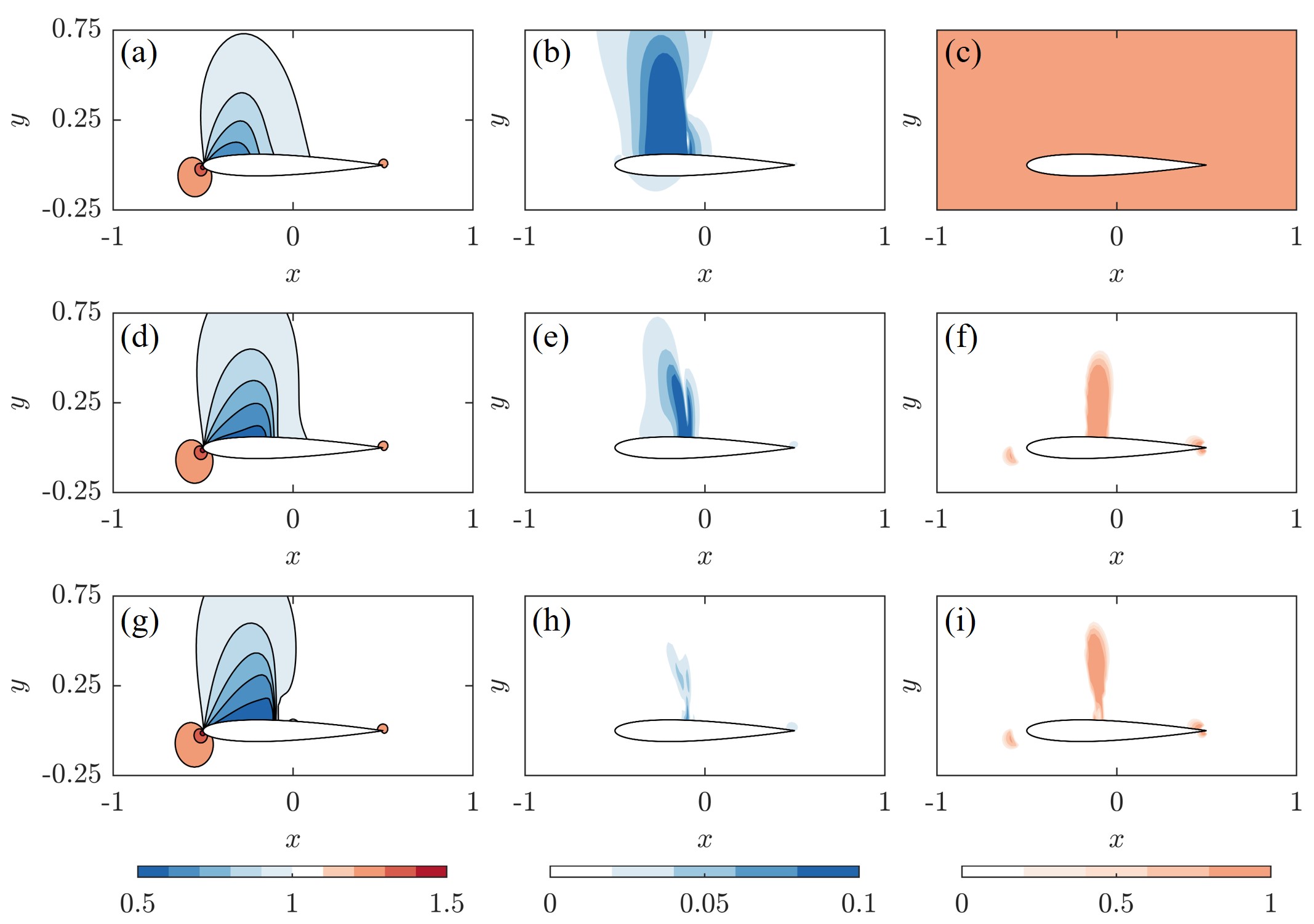}
	\caption{Global pressure solution (left), absolute error (middle), and viscosity distribution (right) obtained from different models for the transonic flow around the NACA0012 airfoil at $\alpha=4^{\circ}$: (a)–(c) MLP with global AV; (d)–(f) MLP with local AV; (g)–(i) DPINN.}
	\label{fig: transonic contourf1}
\end{figure}

In addition to the local airfoil surface pressure profile, we also studied the global pressure solution, absolute error, and viscosity distribution for the flow with shock-waves. The conventional MLP method fails to capture the shock, resulting in large errors around the airfoil (Fig. \ref{fig: transonic contourf0}). The MLP with global AV reduces the error but still fails to capture the shock (Fig. \ref{fig: transonic contourf1}a and \ref{fig: transonic contourf1}b). The MLP with local AV further reduces the error by introducing artificial viscosity only near the shock, resulting in a smooth shock profile (Fig. \ref{fig: transonic contourf1}d and \ref{fig: transonic contourf1}e). In contrast, DPINN captures the sharp shock, with minor errors near the shock (Fig. \ref{fig: transonic contourf1}g and \ref{fig: transonic contourf1}h). As shown in Fig. \ref{fig: transonic contourf1}c, \ref{fig: transonic contourf1}f and \ref{fig: transonic contourf1}i, DPINN restricts viscosity to a smaller region and learns a lower viscosity coefficient of $5.12 \times 10^{-4}$, compared to the MLP with AV methods specified viscosity coefficient of $1.25 \times 10^{-3}$, indicating that the error introduced by the equation modification in DPINN is smaller than MLP with AV methods.

\begin{figure}[H]
	\centering
	\includegraphics[width=0.7525\linewidth]{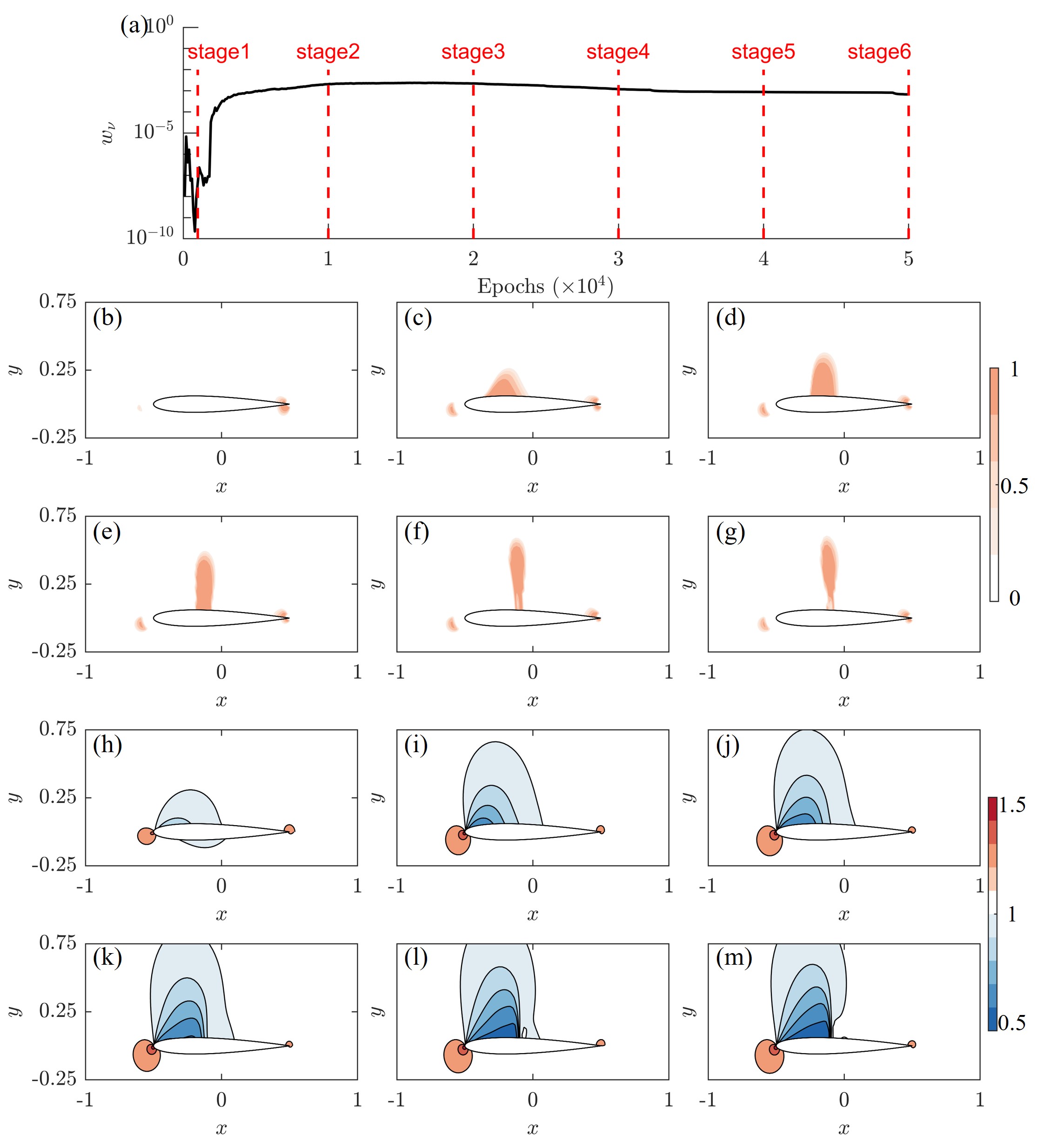}
	\caption{Training evolution: (a) history of the learnable viscosity parameter $w_\nu$; (b–g) artificial viscosity distributions and (h–m) corresponding pressure fields from stages 1 to 6.}
	\label{fig: transonic mu}
\end{figure}

Fig. \ref{fig: transonic mu} presents the training history of $w_\nu$, the evolution of the artificial viscosity distribution, and the corresponding pressure fields. During early training (stage 1), the unstable shock sensor fails to detect or misidentifies shocks. However, this has a negligible impact at this early stage because the initial magnitude of $w_\nu$ is sufficiently small. Due to the rapid convergence of DPINN, the network tends to predict a smooth solution (stage 2), allowing the sensor to detect candidate shock regions. As training progresses (stages 3-6), both the viscosity and the sensor adjust dynamically, ultimately converging to accurate shock solutions.

\begin{figure}[!h]
	\centering
	\includegraphics[width=\linewidth]{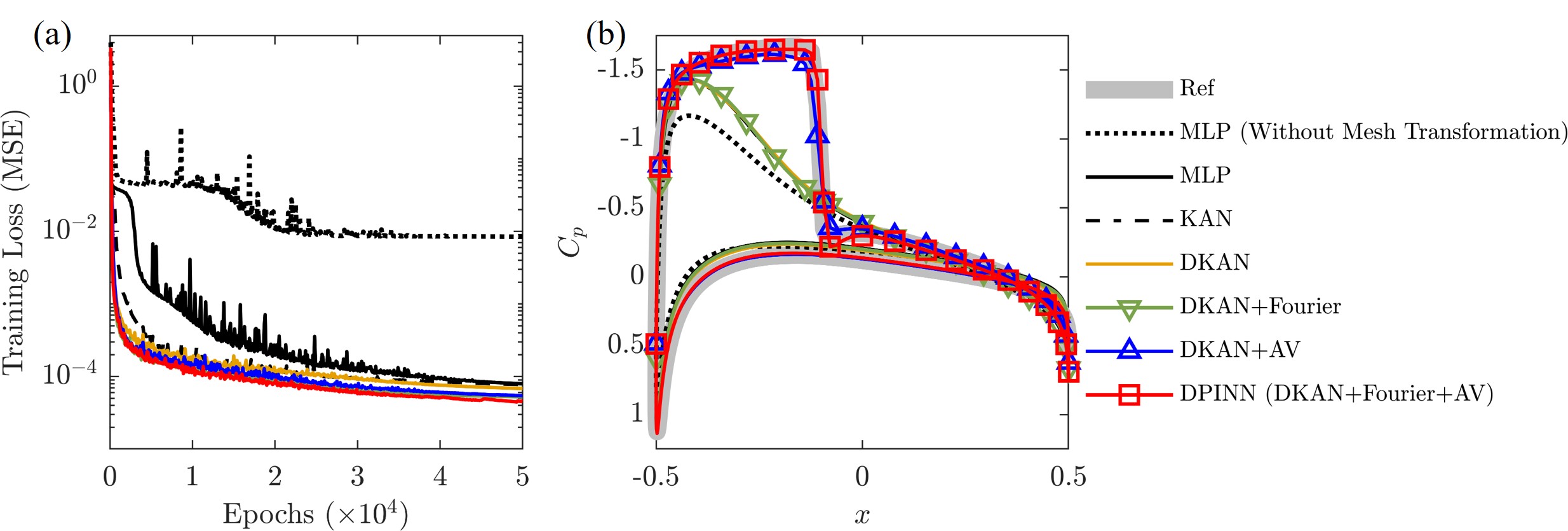}
	\caption{Results of the ablation study for 2D transonic flow over an airfoil. (a) Loss history; (b) pressure coefficient distributions.}
	\label{fig: Ablation_naca0012}
\end{figure}

Fig. \ref{fig: Ablation_naca0012} and Table \ref{T: Ma07 acc} summarize the loss histories and surface pressure distributions for the ablation study. Compared to MLP without mesh transformation (using physical mesh nodes as training points), MLP with mesh transformation reduces the loss by over two orders of magnitude while significantly improving accuracy. Consequently, mesh transformation is indispensable for resolving 2D transonic airfoil flows and is adopted as the default configuration for all evaluated methods. Furthermore, similar to traditional MLP and KAN, both DKAN and DKAN+Fourier produce smooth solutions and fail to resolve shock waves, revealing the inherent limitations of DKAN in high-dimensional scenarios. In contrast, DKAN+AV and DPINN (DKAN+Fourier+AV) successfully capture high-dimensional shock features, with DPINN achieving the highest accuracy and the sharpest solutions.

\subsection{Supersonic flows around an airfoil}
\label{sub:3.3}
Following the transonic case, the method is extended to the supersonic flow around a NACA0012 airfoil at $Ma=1.3$ and $\alpha=0^{\circ}$, with boundary conditions specified in Eq.~\eqref{condition}. Compared to the transonic flow, which features normal shocks on the airfoil surface, the supersonic flow produces an upstream detached bow shock and trailing‐edge oblique shocks. The shock-capturing approach outlined in Section \ref{sub:3.2} is applied with parameters $k_{shock,0}=1$, $k_{shock,1}=4$, $k_{stag,0}=3$, and $k_{stag,1}=5$. Moreover, the network architecture, training strategy, and hyperparameters are identical to those employed in the transonic case.

\begin{figure}[H]
	\centering
	\includegraphics[width=0.985\textwidth]{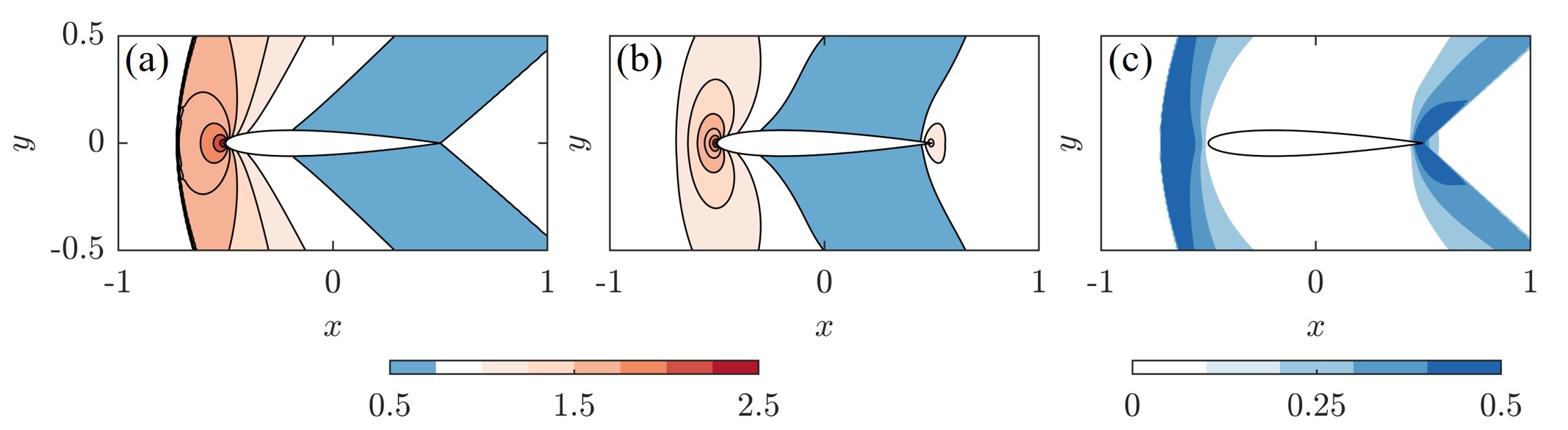}
	\caption{Global pressure field for the supersonic flow around the NACA0012 airfoil at $\alpha=0^{\circ}$. (a) High-fidelity solution; (b) Conventional MLP-based architecture solution; (c) Absolute error.}
	\label{fig: supersonic contourf mlp}
\end{figure}

\begin{figure}[H]
	\centering
	\includegraphics[width=0.985\textwidth]{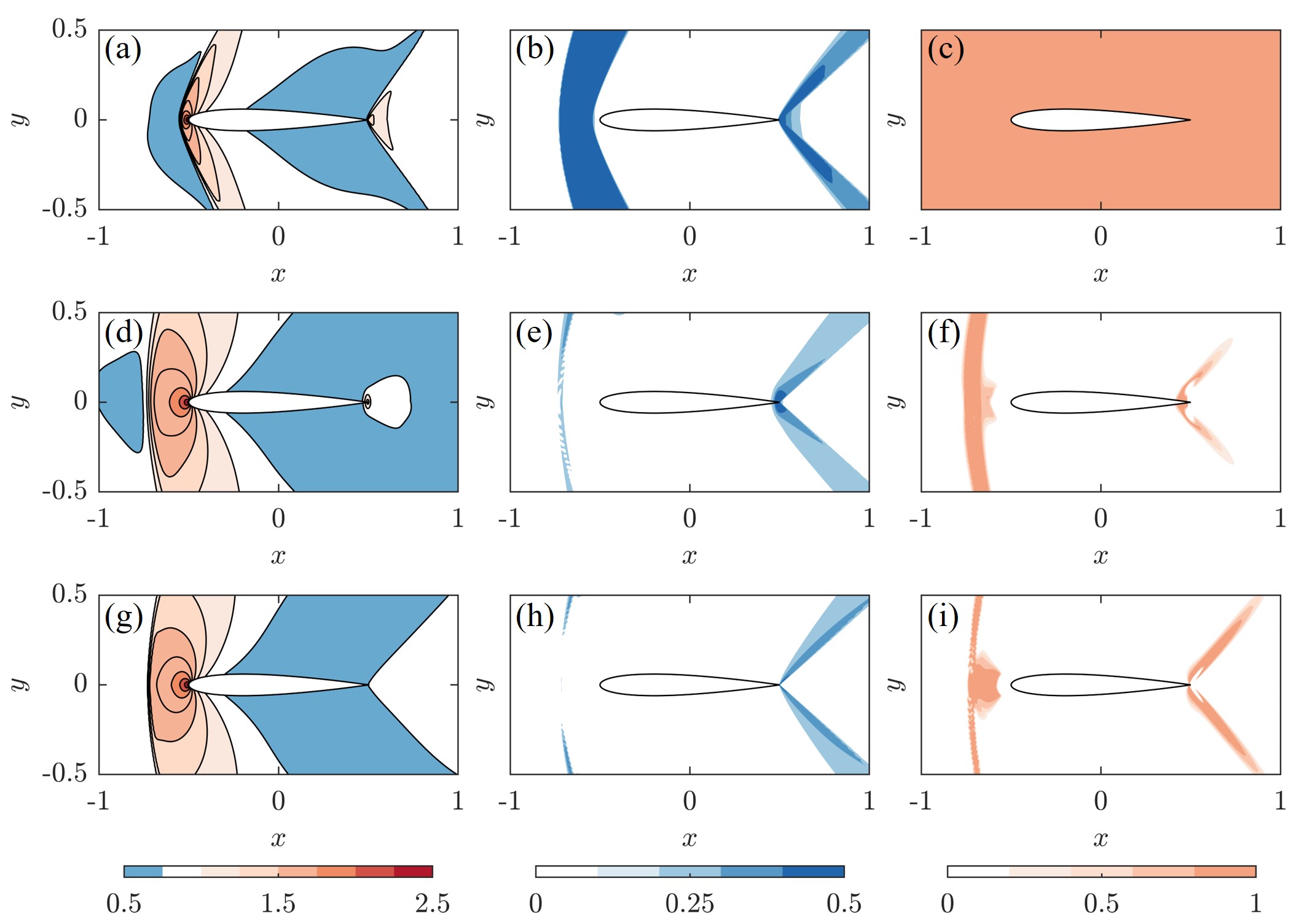}
	\caption{Global pressure solution (left), absolute error (middle), and viscosity distribution (right) obtained from different models for the supersonic flow around the NACA0012 airfoil: (a)–(c) MLP with global AV; (d)–(f) MLP with local AV; (g)–(i) DPINN.}
	\label{fig: supersonic contourf}
\end{figure}

Due to the absence of sharp pressure gradients on the airfoil surface, the performance of these methods is evaluated based on the accuracy of the global pressure solution, and the $R_{front}$ error is therefore not considered, as shown in Table \ref{T: Ma07 acc}. 
The conventional MLP-based architecture fails to capture the leading-edge bow and trailing-edge oblique shock-waves (Fig. \ref{fig: supersonic contourf mlp}). The MLP with global AV method inaccurately predicts the bow shock location and introduces errors in smooth areas, resulting in an even larger $R_2$ error compared to the conventional MLP method. (Fig. \ref{fig: supersonic contourf}a and \ref{fig: supersonic contourf}b). The MLP with local AV produces a smoothed bow shock but fails to capture the sharp oblique shock at the trailing edge (Fig. \ref{fig: supersonic contourf}d and \ref{fig: supersonic contourf}e). DPINN provides the most accurate results, successfully capturing both the detached bow and oblique shocks (Fig. \ref{fig: supersonic contourf}g and \ref{fig: supersonic contourf}h) with error below 5\% (Table \ref{T: Ma07 acc}). As shown in Fig. \ref{fig: supersonic contourf}c, \ref{fig: supersonic contourf}f, and \ref{fig: supersonic contourf}i, DPINN restricts artificial viscosity in a smaller region near the upstream detached bow shock and correctly captures the trailing-edge oblique shock. Additionally, the learned viscosity coefficient of $5.41 \times 10^{-4}$ is smaller than the viscosity coefficient of $1.25 \times 10^{-3}$ specified in the MLP with AV methods, indicating smaller errors from equation modification in DPINN.

\section{Conclusion}
\label{sub:4}
This paper presents a discontinuity-aware PINN method designed to improve the ability of AI-based solvers to resolve PDEs with sharp spatial transitions or fast temporal evolution without prior measured data. It employs an adaptive Fourier-feature embedding with a hybrid spectral correction mechanism to mitigate spectral bias and detect high-frequency discontinuities, a DKAN that generalizes the Kolmogorov representation theorem to the discontinuous regime for the modeling of shock-wave properties, mesh transformation to improve convergence across complex geometries, and learnable local artificial viscosity to stabilize solutions near discontinuities. This method is tested on the Burgers' equation, Riemann problems, and transonic and supersonic airfoil flows, resulting in sharply resolved shocks. These results demonstrate that the proposed DPINN method achieves superior accuracy with fewer parameters in capturing discontinuities compared to other methods.

The efficiency and generalizability of the proposed method can be further improved. In terms of efficiency, the additional artificial viscosity term increases computational overhead arising from second-order derivative evaluations. Theoretically, a sufficiently complex PINN without artificial viscosity can also accurately solve PDEs with discontinuous solutions, as previously shown with architecture modification methods and the inviscid Burgers' equation case in Section \ref{sub:3.1}. However, due to limited computational resources, pure architecture modification methods have so far been limited to 1D cases. Furthermore, the reliance on shock sensors constrains generalizability, as how to accurately and robustly identify shocks from predicted flow fields across diverse scenarios, such as high-dimensional unsteady flows, remains a significant challenge. Therefore, future research will integrate AI-based shock-capturing schemes \cite{fan2025wcns3} and attention mechanisms \cite{song2024loss} to extend DPINN to practical three-dimensional unsteady flows, enhancing its capability to model complex physical phenomena.

\section*{Appendix A: Hyperparameters}
\label{Appendix B}
Detailed hyperparameters for all test cases are provided in Table \ref{T: Hyperparameters}.

\begin{table}[H]
	\centering
	\renewcommand\arraystretch{1.1}
	\begin{tabular}{l|c|c|c|c|c}
		\hline
		Hyperparameter & Burgers & Transonic & Supersonic & Sod & Lax \\
		\hline
		Input dimension & 2 & 2 & 2 & 2 & 2 \\
		Fourier components & 64 & 128 & 128 & 128 & 128 \\
		DKAN hidden layers & 2 & 3 & 3 & 2 & 2 \\
		Neurons/layer & 5 & 12 & 12 & 12 & 12 \\
		Output dimension & 1 & 4 & 4 & 3 & 3 \\
		\hline
		$k_{\text{shock},0},k_{\text{shock},1}$ & 0.015,1 & 0.5,5 & 1,4 & 0.5,1 & 0.5,1 \\
		$k_{\text{stag},0},k_{\text{stag},1}$ & - & 3,4 & 3,5 & - & - \\
		$\lambda_{\text{pde}},\lambda_{\text{ic}},\lambda_{\text{bc}}$ & 1,1,1 & 20000, -, 1 & 20000, -, 1 & 1,1,1 & 1,1,1 \\
		Optimizer & AdamW & SOAP & SOAP & AdamW & AdamW \\
		Epochs & 50000 & 50000 & 50000 & 50000 & 50000 \\
		Batch size & 20000 & 20000 & 20000 & 20000 & 20000 \\
		Learning rate & 0.001 & 0.003 & 0.003 & 0.001 & 0.001 \\
		\hline
	\end{tabular}
	\caption{Summary of hyperparameters.}
	\label{T: Hyperparameters}
\end{table}

\section*{Appendix B: Hyperparameter sensitivity}
\label{Appendix C}
For 2D transonic flows over an airfoil, the shock sensor is governed by four parameters (baseline): $k_{\text{shock},0}=0.5$ and $k_{\text{shock},1}=5$ control the detection threshold and shock sensor steepness, respectively, while $k_{\text{stag},0}=3$ and $k_{\text{stag},1}=4$ define the stagnation detection threshold and switching function steepness to prevent the misidentification of stagnation regions as shocks.

To evaluate hyperparameter sensitivity, 8 experiments are performed where each parameter is scaled by a factor of 10 and 0.1 relative to the baseline. Table \ref{T: sensor_sensitivity} summarizes the relative errors in airfoil surface pressure across these configurations. The $k_{\text{shock},0} \times 10$ case yields the highest error, indicating that the shock detection threshold is the most sensitive parameter. Conversely, $k_{\text{shock},1}$ has minimal impact, while the stagnation point parameters $k_{\text{stag},0}$ and $k_{\text{stag},1}$, which prevent the misidentification of stagnation regions as shocks, exert an intermediate impact.

\begin{figure}[H]
	\centering
	\includegraphics{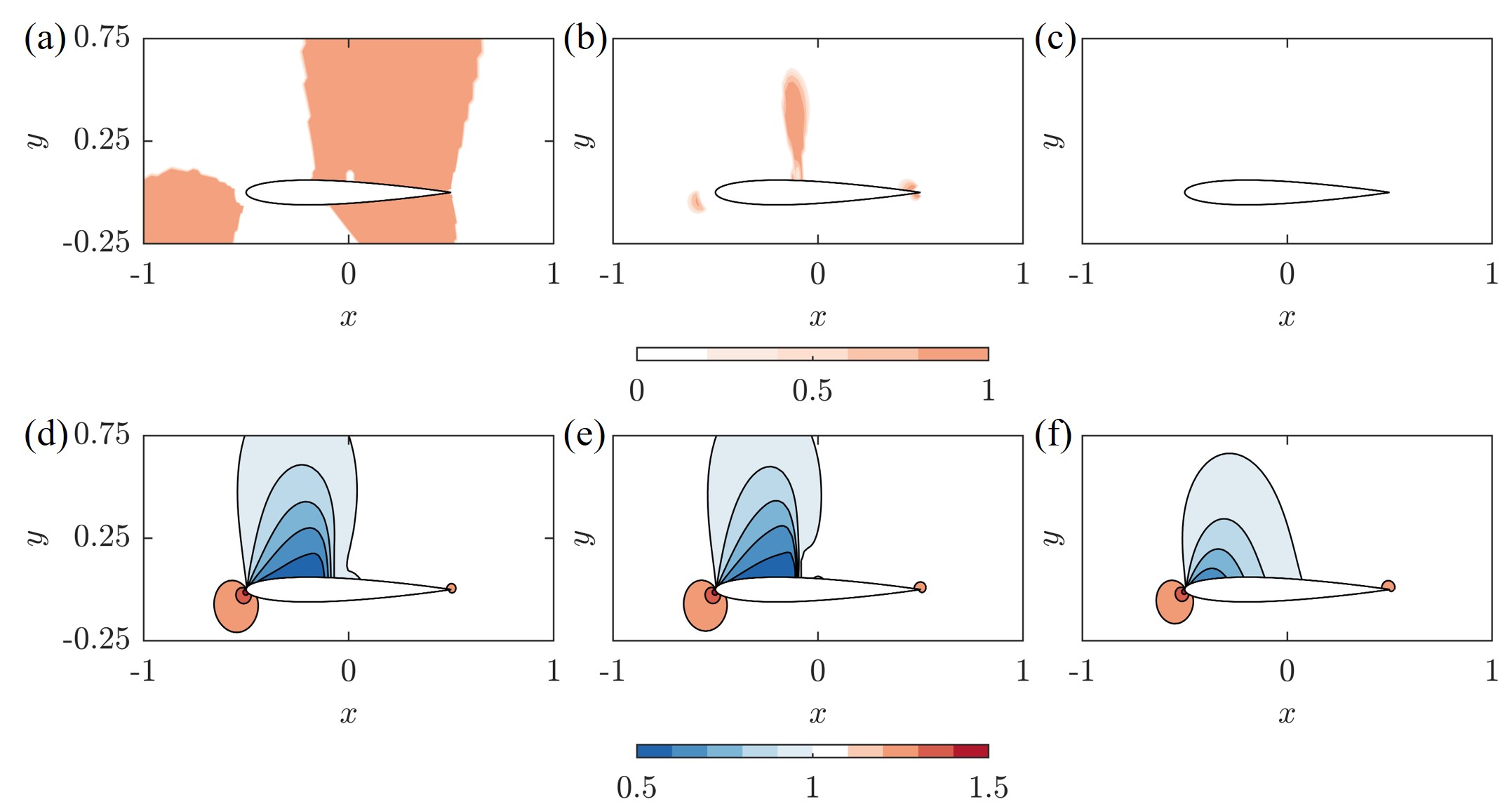}
	\caption{(a–c) Artificial viscosity distributions and (d–f) corresponding pressure fields for $k_{\text{shock},0} \times 0.1$, baseline, and $k_{\text{shock},0} \times 10$ cases, respectively.}
	\label{fig: transonic prameter}
\end{figure}

Fig. \ref{fig: transonic prameter} compares the shock sensor distributions and flow solutions for $k_{\text{shock},0} \times 0.1$, baseline, and $k_{\text{shock},0} \times 10$ cases. When $k_{\text{shock},0}$ is excessively small, unnecessary artificial viscosity is introduced into shock-free regions, slightly dissipating the shock. Conversely, an excessive $k_{\text{shock},0}$ leads to an overly strict detection criterion, preventing shock detection and resulting in a smooth solution.

\begin{table}[htbp]
	\centering
	\begin{tabular}{@{}l | c c c@{}} 
		\hline
		\multirow{2}{*}[-0.5ex]{Group} & 
		\multicolumn{3}{c}{Relative Error (\%)} \\
		
		\cline{2-4} 
		& $R_1$ & $R_2$ & $R_\infty$ \\
		\hline
		
		\rowcolor{gray!20} Baseline & 3.17 & 3.80 & 0.39 \\
		\hline
		$k_{\text{shock},0} \times 0.1$ & 6.77 & 9.14 & 1.97 \\
		$k_{\text{shock},0} \times 10$  & 28.17 & 30.11 & 19.40 \\
		\hline
		$k_{\text{shock},1} \times 0.1$ & 3.89 & 5.01 & 0.96 \\
		$k_{\text{shock},1} \times 10$  & 3.32 & 4.33 & 0.47 \\
		\hline
		$k_{\text{stag},0} \times 0.1$  & 4.75 & 6.24 & 1.50 \\
		$k_{\text{stag},0} \times 10$   & 7.45 & 8.24 & 2.12 \\
		\hline
		$k_{\text{stag},1} \times 0.1$  & 5.60 & 6.51 & 1.73 \\
		$k_{\text{stag},1} \times 10$   & 3.73 & 4.29 & 0.78 \\
		\hline
	\end{tabular}
	\caption{Results of the parameter sensitivity analysis for the shock sensor.}
	\label{T: sensor_sensitivity}
\end{table}

\section*{Appendix C: One-dimensional Riemann problems}
\label{Appendix A}

\begin{figure}[H]
	\centering
	\includegraphics[width=0.675\linewidth]{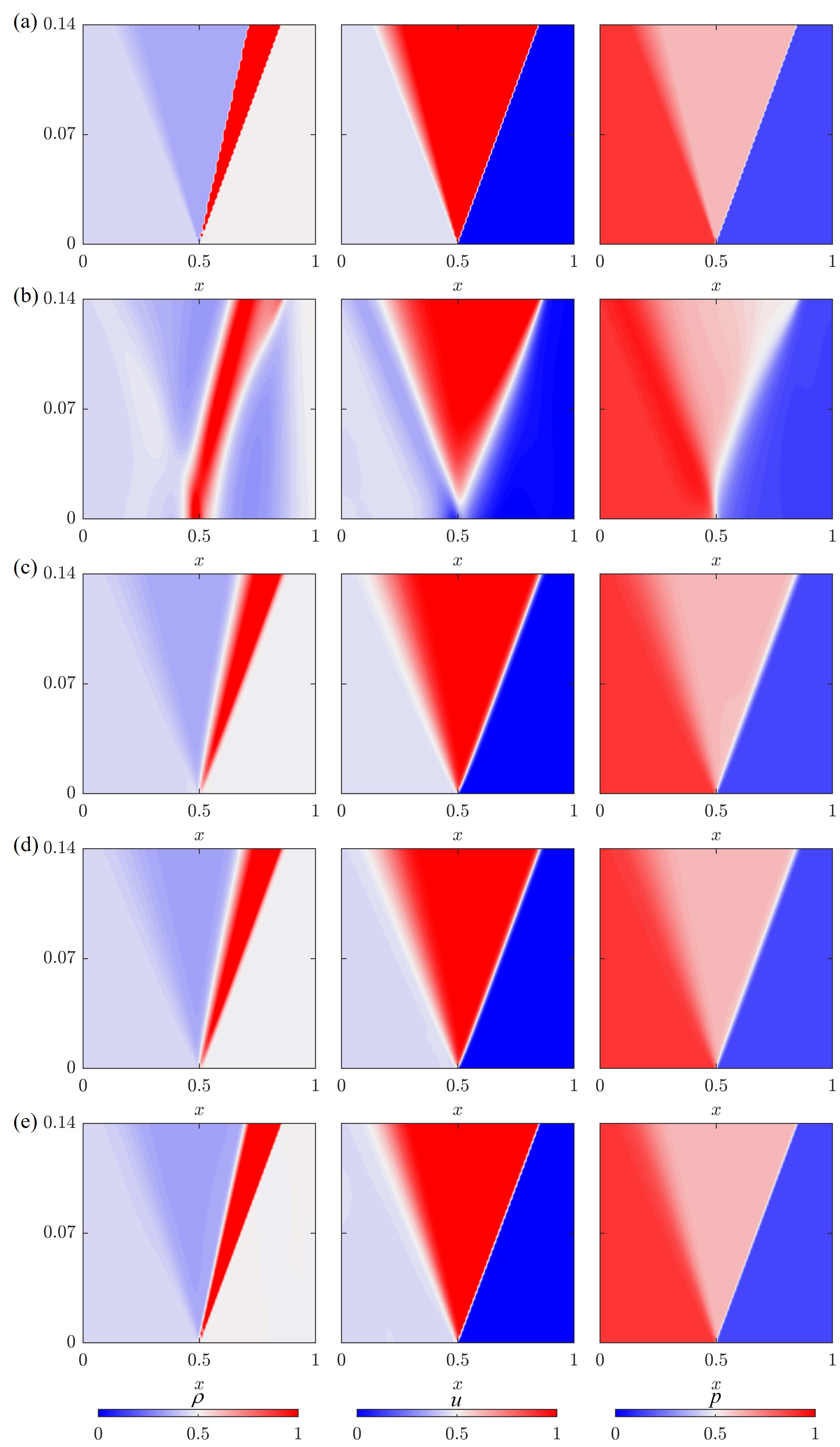}
	\caption{Sod shock tube problem: velocity, pressure, and density contours (left to right) obtained from (a) reference solution; (b) MLP; (c) MLP with global AV; (d) MLP with local AV; and (e) DPINN.}
	\label{fig: SST_contourf}
\end{figure}

\begin{figure}[H]
	\centering
	\includegraphics[width=0.675\linewidth]{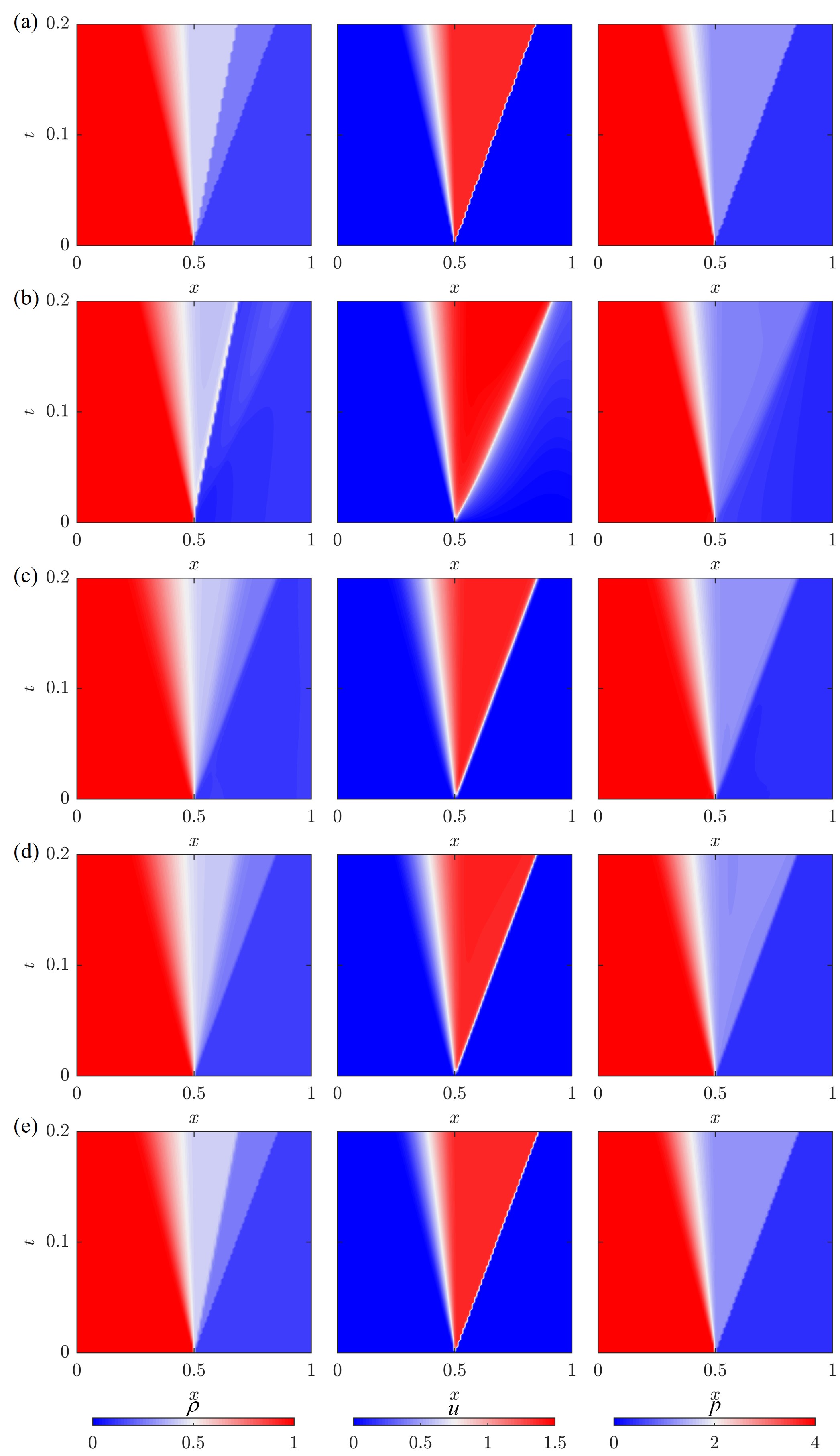}
	\caption{Lax shock tube problem: velocity, pressure, and density contours (left to right) obtained from (a) reference solution; (b) MLP; (c) MLP with global AV; (d) MLP with local AV; and (e) DPINN.}
	\label{fig: Lax_contourf}
\end{figure}

\section*{Declaration of competing interest}
The authors declare that they have no known competing financial interests or personal relationships that could have appeared to influence the work reported in this paper.
\section*{Data availability}
Data will be made available on request.
\section*{Acknowledgments}
The authors would like to acknowledge the support from the National Natural Science Foundation of China (Grant Nos. 12202059).

\bibliographystyle{elsarticle-num-names} 
\bibliography{reference2}
\biboptions{numbers,sort&compress}






\end{document}